\documentclass[twoside,twocolumn,9pt]{article}
\usepackage{extsizes}
\usepackage[super,sort&compress,comma]{natbib} 
\usepackage[version=3]{mhchem}
\usepackage[left=1.5cm, right=1.5cm, top=1.785cm, bottom=2.0cm]{geometry}
\usepackage{balance}
\usepackage{}
\usepackage{mathptmx}
\usepackage{sectsty}
\usepackage{graphicx} 
\usepackage{placeins}
\usepackage{lastpage}
\usepackage[format=plain,justification=justified,singlelinecheck=false,font={stretch=1.125,small,sf},labelfont=bf,labelsep=space]{caption}
\usepackage{float}
\usepackage{fancyhdr}
\usepackage{tikz}
\usepackage{fnpos}
\usepackage[english]{babel}
\addto{\captionsenglish}{%
  \renewcommand{\refname}{Notes and References}
}

\usepackage{appendix}  
\usepackage{enumitem}

\usepackage{array}
\usepackage{droidsans}
\usepackage{charter}
\usepackage{svg}
\usepackage[utf8]{inputenc}
\usepackage{graphicx}
\usepackage{wrapfig}
\usepackage{placeins}
\usepackage{svg}
\usepackage{float}
\usepackage{booktabs}
\usepackage[T1]{fontenc}
\usepackage{setspace}
\usepackage{adjustbox}
\usepackage{multirow}
\usepackage{multicol}
\usepackage[compact]{titlesec}
\usepackage[hidelinks]{hyperref}
\usepackage{siunitx}
\usepackage{wasysym}
\usepackage{titlesec}
\titleformat{\subsection}[runin]{\normalfont\bfseries}{\thesubsection.}{1em}{}[.]
\usepackage[nonumberlist,style=super,acronym,toc]{glossaries}
\setacronymstyle{long-short}
\newglossary[slg]{symbolslist}{syi}{syg}{List of Symbols}

\makeglossaries

\newacronym[description={This is your acronym description}]{abc}{ABC}{\MakeLowercase{abc}}
\newacronym[description={Methanol}]{meOH}{MeOH}{\MakeLowercase{methanol}}
\newacronym[description={Silver nitrate}]{agno3}{AgNO$_3$}{\MakeLowercase{silver nitrate}}
\newacronym{sem}{SEM}{Scanning electron microscopy}
\newacronym[description={Strontium sulfide}]{srs}{SrS}{\MakeLowercase{Strontium sulfide}}
\newacronym[description={Aluminium oxide}]{al2o3}{Al$_2$O$_3$}{\MakeLowercase{aluminium oxide}}
\newacronym[description={Strontium iodide}]{sri2}{SrI$_2$}{Strontium iodide}
\newacronym[description={scanning transmission electron microscopy}]{stem}{STEM}{\MakeLowercase{scanning transmission electron microscopy}}
\newacronym[description={Nuclear magnetic resonance spectroscopy}]{nmr}{NMR}{nuclear magnetic resonance spectroscopy}
\newacronym[description={Transmission electron microscopy}]{tem}{TEM}{\MakeLowercase{transmission electron microscopy}}
\newacronym[description={Fourier-transform infrared spectroscopy}]{ftir}{FT-IR}{Fourier-transform infrared spectroscopy}
\newacronym[description={Density functional theory}]{dft}{DFT}{\MakeLowercase{density functional theory}}
\newacronym[description={X-ray photoelectron spectroscopy}]{xps}{XPS}{X-ray photoelectron spectroscopy}
\newacronym{xrf}{XRF}{X-ray fluorescence}
\newacronym{pl}{PL}{photoluminescence}
\newacronym{trpl}{TRPL}{time-resolved photoluminescence}
\newacronym[description={Energy-dispersive X-ray spectroscopy}]{eds}{EDS}{\MakeLowercase{energy-dispersive X-ray spectroscopy}}
\newacronym[description={Spectroscopic ellipsometry}]{se}{SE}{\MakeLowercase{spectroscopic ellipsometry}}
\newacronym[description={Electrodepostion}]{ecd}{ECD}{\MakeLowercase{electrochemical deposition}}
\newacronym[description={Cyclic Amperometry}]{ca}{CA}{\MakeLowercase{cyclic amperometry}}
\newacronym[description={Dynamic light scattering}]{dls}{DLS}{\MakeLowercase{dynamic light scattering}}
\newacronym[description={Electroless deposition}]{eld}{ELD}{\MakeLowercase{electroless deposition}}
\newacronym[description={Nanoparticle}]{np}{NP}{\MakeLowercase{nanoparticle}}
\newacronym[description={Nanocrystal}]{nc}{NC}{\MakeLowercase{nanocrystal}}
\newacronym[description={Quantum dot}]{qds}{QD}{\MakeLowercase{quantum dot}}
\newacronym{xrd}{XRD}{X-ray diffraction}
\newacronym{cits}{CuInSnS$_4$}{copper indium tin sulfide}
\newacronym[description={Atomic layer deposition}]{ald}{ALD}{\MakeLowercase{atomic layer deposition}}
\newacronym[description={Trimethylaluminium}]{tma}{TMA}{Trimethylaluminium}
\newacronym[description={Alkaline Earth Metal Sulfides}]{aes2}{(A$_E$)S$_2$}{\MakeLowercase{alkaline earth metal sulfides}}
\newacronym{dmtu}{DMTU}{N,N'-dimethylthiourea}
\newacronym[description={Relative centrifugal force}]{rcf}{RCF}{\MakeLowercase{relative centrifugal force}}
\newacronym{etoh}{EtOH}{Ethanol}
\newacronym[description={Scanning near-field optical microscopy}]{snom}{SNOM}{\MakeLowercase{scanning near-field optical microscopy}}
\newacronym[description={Oleic acid}]{oa}{OA}{\MakeLowercase{oleic acid}}
\newacronym[description={Kelvin probe force microscopy}]{kpfm}{KPFM}{\MakeLowercase{kelvin probe force microscopy}}
\newacronym[description={Oleylamine}]{ola}{OLA}{oleylamine}
\newacronym[description={Titanium(IV) isopropoxide}]{ttip}{TTIP}{\MakeLowercase{titanium(IV) isopropoxide}}
\newacronym[description={Barium(0)}]{ba}{Ba}{Barium(0)}
\newacronym[description={Barium}]{ba1}{Ba}{barium}
\newacronym[description={Barium oxide}]{bao}{BaO}{barium oxide}
\newacronym[description={Titanium(0)}]{ti}{Ti}{titanium(0)}
\newacronym[description={Titanium}]{ti1}{Ti}{titanium}
\newacronym[description={Titanium dioxide}]{tio}{TiO$_2$}{titanium dioxide}
\newacronym[description={Acetone}]{acoh}{AcOH}{acetone}
\newacronym[description={Barium titanate}]{bto}{BaTiO$_3$}{Barium titanate}
\newacronym[description={Decanoic acid}]{da}{DA}{decanoic acid}
\newacronym[description={Benzyl alcohol}]{bzoh}{BzOH}{benzyl alcohol}
\newacronym[description={Oleyl alcohol}]{oloh}{OLOH}{oleyl alcohol}

\definecolor{cream}{RGB}{222,217,201}

\begin{document}

\pagestyle{fancy}
\thispagestyle{plain}
\fancypagestyle{plain}{
\renewcommand{\headrulewidth}{0pt}
}

\makeFNbottom
\makeatletter
\renewcommand\LARGE{\@setfontsize\LARGE{15pt}{17}}
\renewcommand\Large{\@setfontsize\Large{12pt}{14}}
\renewcommand\large{\@setfontsize\large{10pt}{12}}
\renewcommand\footnotesize{\@setfontsize\footnotesize{7pt}{10}}
\makeatother

\renewcommand{\thefootnote}{\fnsymbol{footnote}}
\renewcommand\footnoterule{\vspace*{1pt}%
\color{cream}\hrule width 3.5in height 0.4pt \color{black}\vspace*{5pt}} 
\setcounter{secnumdepth}{5}

\makeatletter 
\renewcommand\@biblabel[1]{#1}            
\renewcommand\@makefntext[1]%
{\noindent\makebox[0pt][r]{\@thefnmark\,}#1}
\makeatother 
\renewcommand{\figurename}{\small{Fig.}~}
\sectionfont{\sffamily\Large}
\subsectionfont{\normalsize}
\subsubsectionfont{\bf}
\setstretch{1.125} 
\setlength{\skip\footins}{0.8cm}
\setlength{\footnotesep}{0.25cm}
\setlength{\jot}{10pt}
\titlespacing*{\section}{0pt}{1pt}{2pt}
\titlespacing*{\subsection}{0pt}{1pt}{2pt}


\fancyfoot{}
\fancyfoot[RO]{\footnotesize{\sffamily{1--\pageref{LastPage} ~\textbar  \hspace{2pt}\thepage}}}
\fancyfoot[LE]{\footnotesize{\sffamily{\thepage~\textbar 1--\pageref{LastPage}}}}
\fancyhead{}
\renewcommand{\headrulewidth}{0pt} 
\renewcommand{\footrulewidth}{0pt}
\setlength{\arrayrulewidth}{1pt}
\setlength{\columnsep}{6.5mm}
\setlength\bibsep{1pt}

\makeatletter 
\newlength{\figrulesep} 
\setlength{\figrulesep}{0.5\textfloatsep} 

\newcommand{\topfigrule}{\vspace*{-1pt}%
\noindent{\color{cream}\rule[-\figrulesep]{\columnwidth}{1.5pt}} }

\newcommand{\botfigrule}{\vspace*{-2pt}%
\noindent{\color{cream}\rule[\figrulesep]{\columnwidth}{1.5pt}} }

\newcommand{\dblfigrule}{\vspace*{-1pt}%
\noindent{\color{cream}\rule[-\figrulesep]{\textwidth}{1.5pt}} }

\makeatother

\twocolumn[
  \begin{@twocolumnfalse}

\vspace{1em}
\sffamily
\begin{tabular}{m{0cm} p{17cm} }

 & \noindent\LARGE{\textbf{When Cubic Is Not Isotropic: Phonon–Exciton Decoupling in CuInSnS$_4$ single crystals}}\\
\vspace{0.3cm} & \vspace{0.3cm} \\

 & \noindent\large{Lara Kim Linke $^{a,g}$, Yvonne Tomm $^{b}$, Xinyun Liu $^{c}$, Galina Gurieva $^{b}$, Daniel M. Többens$^{b}$, Pardis Adams $^{d}$, Michel Calame $^{a,e,f}$, Ryan W. Crisp $^{g}$, Jessica Boland $^{c}$, Seán Kavanagh $^{h}$, Susan Schorr $^{b,i}$, Mirjana Dimitrievska $^{a,d*}$ }

\end{tabular}

 \end{@twocolumnfalse} \vspace{0.6cm}

  ]

\renewcommand*\rmdefault{bch}\normalfont\upshape
\rmfamily
\section*{}
\vspace{-1cm}


\footnotetext{\textit{$^{a}$ Transport at Nanoscale Interfaces Laboratory, Swiss Federal Laboratories for Material Science and Technology (EMPA), Ueberlandstrasse 129, Dübendorf 8600, Switzerland; E-mail: \textbf{mirjana.dimitrievska@empa.ch}}}

\footnotetext{\textit{$^{b}$ Helmholtz-Zentrum Berlin für Materialien und Energie, Department Structure and Dynamics of Energy Materials, Hahn-Meitner-Platz 1, 14109 Berlin, Germany}}

\footnotetext{\textit{$^{c}$Photon Science Institute and Department of Materials, University of Manchester, Manchester M13 9PL, United Kingdom}}

\footnotetext{\textit{$^{d}$Department of Chemistry, University of Zurich, Winterthurerstrasse 190, Zurich 8057, Switzerland}}

\footnotetext{\textit{$^{e}$ Department of Physics, University of Basel, 4056 Basel, Switzerland; Swiss Nanoscience Institute, University of Basel, 4056 Basel, Switzerland}}

\footnotetext{\textit{$^{f}$ Swiss Nanoscience Institute, University of Basel, 4056 Basel, Switzerland}}

\footnotetext{\textit{$^{g}$ Chemistry of Thin Film Materials, Department of Chemistry and Pharmacy, Friedrich-Alexander-Universität Erlangen-Nürnberg, Cauerstr. 3, 91058 Erlangen, Germany;}}

\footnotetext{\textit{$^{h}$ Yusuf Hamied Department of Chemistry, University of Cambridge, Cambridge, Lensfield Road, Cambridge, CB2 1EW, United Kingdom; }}

\footnotetext{\textit{$^{i}$ Freie Universität Berlin, Malteserstraße 74‑100, 12249 Berlin, Germany}}

\vspace{1.6 em}
\hrule
\vspace{1em}
\begin{abstract}
Atomic-scale disorder can create hidden optical anisotropy even in crystals that are structurally cubic on average. Here, we show that CuInSnS$_4$ single crystals host locally symmetry-broken environments arising from intrinsic In/Sn cation disorder, which affect vibrational and excitonic properties in markedly different ways. Combining polarization- and temperature-dependent Raman spectroscopy, infrared near-field microscopy, steady-state and time-resolved photoluminescence, and first-principles calculations, we find that phonons remain largely symmetry-averaged and locally homogeneous on the nanoscale. In contrast, photoluminescence reveals a lower-energy band-tail emission with pronounced polarization anisotropy following a well-defined angular symmetry, highlighting the strong sensitivity of excitonic states to local symmetry breaking. This phonon--exciton decoupling reveals that intrinsic disorder can localize excitons while preserving vibrational coherence and dielectric homogeneity, thereby opening new opportunities for polarization-sensitive light sources, anisotropic photodetectors, and exciton-based optical functionalities even in nominally cubic multinary semiconductors.

\end{abstract}
\vspace{0.5em}
\begin{center}
  \includegraphics[width=0.5\textwidth]{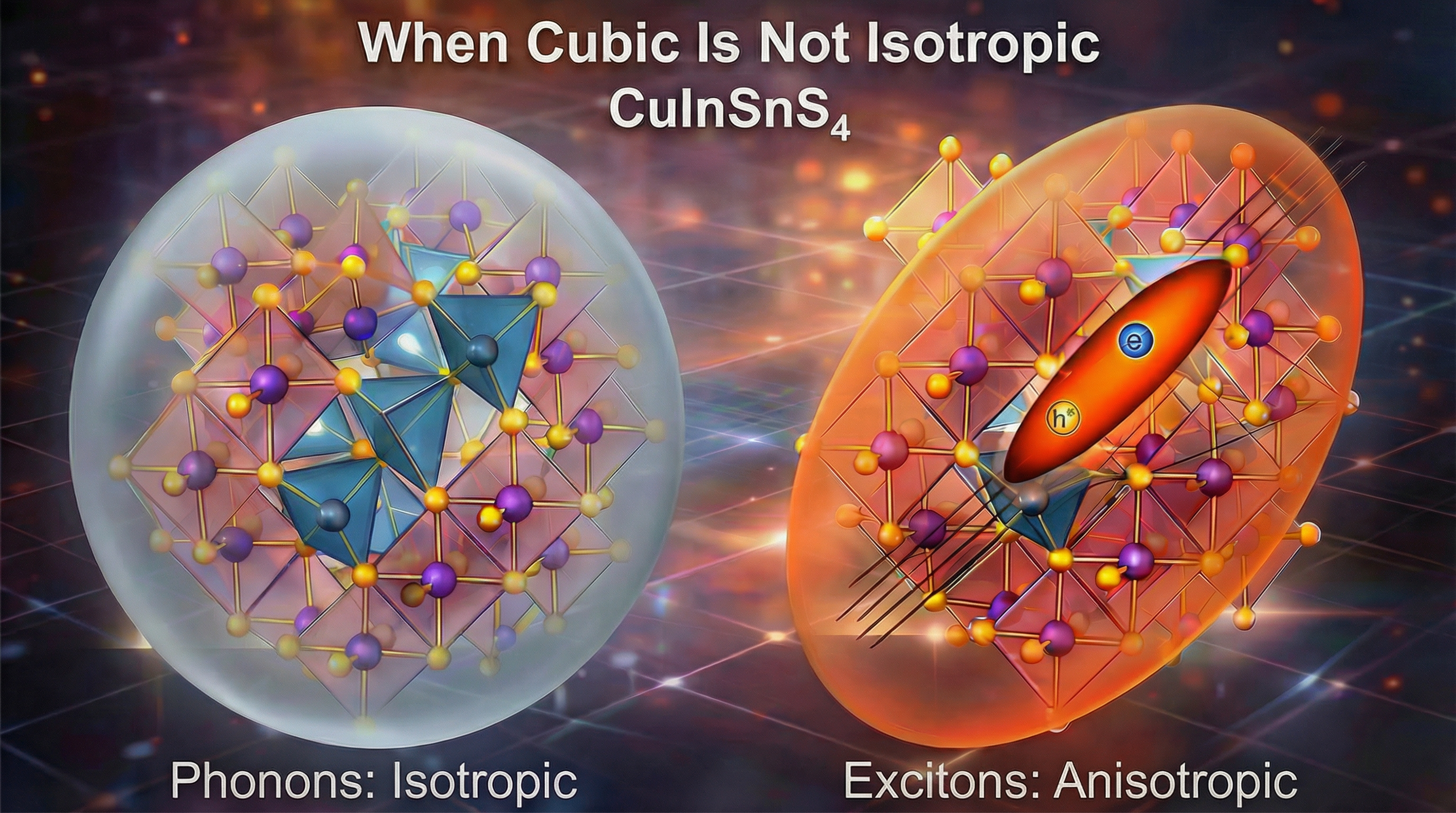}
\end{center}

\vspace{0.5em}
\hrule

\section{Introduction}

Atomic-scale disorder is becoming increasingly recognized as an active “design parameter” in optical semiconductors rather than a structural imperfection.\cite{zeng2026control, baranovskii2022energy,schnepf2020utilizing} Even subtle cation-site mixing, alloy composition fluctuations, or local symmetry breaking can reshape the electronic density of states near the band edges, producing Urbach tails and spatial band-gap fluctuations that broaden absorption and photoluminescence (PL), modify carrier mobilities, and shift recombination pathways toward (or away from) radiative channels.\cite{piccardo2017localization, wong2020impact} In parallel, disorder can alter exciton behavior where band-edge states may fragment into localized states, develop sizable Stokes shifts, and couple more strongly to phonons.\cite{vlaming2011disorder, martin1999exciton} These effects can modify carrier lifetimes and, in locally low-symmetry environments, give rise to emergent polarization anisotropies.\cite{ni2017real, vlaming2011disorder}

However, disorder does not affect all material properties in the same way.\cite{nicolson_interplay_2023,roth_tuning_2023,liga_mixedcation_2023} While phonons often reflect the average crystal symmetry, electronic and excitonic states can be sensitive to local symmetry breaking on much shorter length scales (within a unit cell). As a result, the same disordered lattice can exhibit markedly different symmetry signatures depending on whether it is probed through phonons or through electronic excitations. Understanding when disorder is detrimental (e.g., nonradiative trapping, hopping-limited transport) versus beneficial (e.g., localization-protected emission, defect tolerance, engineered broad-band emission) is therefore central to rationally tuning optoelectronic performance.\cite{mosquera2025multifaceted, schnepf2020utilizing}

In III-nitride light emitters, disorder is a textbook example of “useful imperfection.”\cite{weisbuch2020disorder} In InGaN alloys, compositional fluctuations and In-rich local motifs create nanoscale potential landscapes that localize holes/excitons, helping maintain high internal quantum efficiency even in the presence of high threading-dislocation densities.\cite{weisbuch2020disorder} This defect-insensitive emission has been directly connected to carrier localization on very short length scales and to excitonic recombination dynamics in In-containing (Al,In,Ga)N systems.\cite{lottigier2023investigation, chichibu2006origin} In this family, the same disorder physics that can impede long-range transport can also be exploited to tune emission color and stabilize radiative recombination against extended defects.\cite{chichibu2006origin} This illustrates why quantitatively linking microscopic disorder to optical properties, carrier capture, and recombination kinetics is essential for rational materials design.

On the other hand, halide perovskites offer a complementary and highly dynamic disorder paradigm, in which strong anharmonic lattice motion and local structural fluctuations coexist with remarkably sharp absorption edges and low Urbach energies in high-quality films.\cite{gehrmann2019dynamic} Precise absorption studies on halide perovskites reveal steep band-edge onsets with high photovoltages, while theory and experiment show that temperature-driven lattice dynamics continuously reshape the disorder potential on short length scales.\cite{gehrmann2019dynamic, ledinsky2019temperature,frohna_nanoscale_2022} In multinary chalcogenides, disorder becomes largely configurational and often cation-sublattice–driven, with direct consequences for optical and transport properties. In kesterites, nanoscale band-gap fluctuations and defect complexes produce substantial band tailing that limits device voltages.\cite{larsen2020band} More broadly across multinary tetrahedrally bonded semiconductors, extended antisites and entropy-stabilized disorder are now understood as key levers that can either enable defect tolerance or enforce localization and recombination losses, depending on how (and where) the disorder is expressed. \cite{baranowski2016review}

In this work, we show that CuInSnS$_4$, a semiconductor especially promising for optoelectronic applications and photocatalytic CO$_2$ reduction,\cite{chai2023metal,el2023effect} belongs to a material class in which an apparently high-symmetry cubic structure hides locally symmetry-broken environments that directly affect the optical response. Because In$^{3+}$ and Sn$^{4+}$ ordering is difficult to resolve by conventional diffraction measurements, we combine polarization- and temperature-dependent vibrational (Raman and infrared near-field (IR-SNOM)) spectroscopy with steady-state and time-resolved (TR) photoluminescence (PL) to disentangle vibrational and excitonic responses to disorder. Phonon modes reflect a symmetry-averaged lattice response consistent with the cubic structure. On the other hand, the PL consists of two overlapping emission channels, indicating that electronic states are sensitive to local symmetry variations associated with cation disorder. Polarization-resolved PL reveals pronounced anisotropy in the lower-energy emission, whereas the higher-energy contribution remains nearly isotropic, consistent with band-to-band recombination. This contrast is schematically illustrated in Figure~1, where the vibrational response remains effectively isotropic despite local disorder, while the excitonic response becomes anisotropic due to the sensitivity of electronic states to locally symmetry-broken environments. Together, these results establish CuInSnS$_4$ as a model system for disentangling how intrinsic cation disorder shapes vibrational and excitonic responses in multinary chalcogenides, paving the way for disorder engineering of carrier localization, polarization-dependent emission, and radiative efficiency in optoelectronic and photocatalytic devices.

\begin{figure}[h]
    \centering
    \includegraphics[width=\columnwidth]{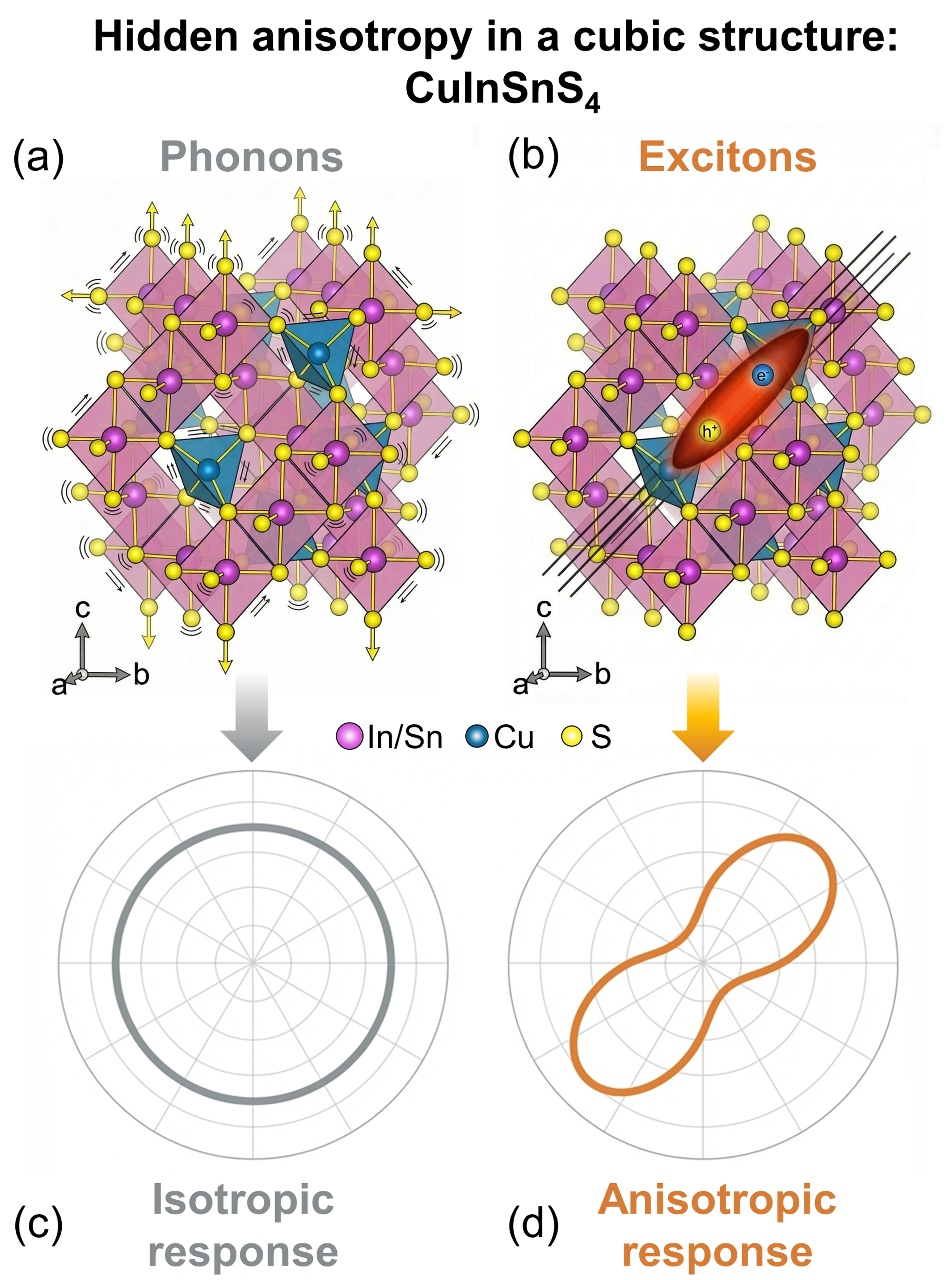}
   \caption{Hidden anisotropy in a cubic structure of CuInSnS$_4$. Although the average crystal structure is cubic, phonons and excitons sense disorder in different ways. (a) Phonon modes reflect a symmetry-averaged lattice response consistent with the cubic structure. (b) In contrast, localized excitons are sensitive to local symmetry breaking associated with cation disorder and exhibit a preferred orientation. (c,d) The corresponding schematics illustrate an effectively isotropic vibrational response (c) and a pronounced anisotropic excitonic response (d) in CuInSnS$_4$.}
    \label{crystal}
\end{figure}

\section{Results}

\subsection{Structural characterization and lattice dynamics}

We begin with the structural characterization of a CuInSnS$_4$ single crystal and compare ordered and disordered structural models to understand their effects on lattice dynamics and phonon behavior.

Figure~\ref{crystal} shows an optical image of the CuInSnS$_4$ single crystal used throughout this study. The crystal exhibits well-defined facets and millimeter-scale dimensions. The use of a bulk single crystal enables a direct comparison between our spectroscopic measurements and first-principles lattice-dynamical calculations without additional effects arising from grain boundaries or preferred orientation effects.

The chemical composition of the CuInSnS$_4$ crystal was determined by energy-dispersive X-ray spectroscopy (EDS), yielding Cu 15 at.\%, In 15 at.\%, Sn 15 at.\%, and S 55 at.\%. Within the typical EDS uncertainty of approximately $\pm$2--3 at.\%, the measured composition is consistent with the intended Cu:In:Sn:S = 1:1:1:4 quaternary stoichiometry and shows indistinguishable In and Sn concentrations, reflecting their chemical similarity and a high tendency for cation mixing on crystallographically equivalent sites.

Synchrotron XRD measurements performed on crushed single crystals were analyzed using Le Bail refinement.\cite{le2005whole,dimitrievska2016role} As shown in Figure~\ref{crystal}, the diffraction pattern is well described by a cubic spinel-type structure with space group Fd$\bar{3}$m and a refined lattice parameter of $a = 10.495 \pm 0.002$~\AA, which is in good agreement with previously reported values.\cite{ohachi1977growth}  Within this average structural model, sulfur atoms form a cubic close-packed anion framework, Cu$^{+}$ occupies tetrahedral sites, and In$^{3+}$ and Sn$^{4+}$ reside on octahedral sites. The Wyckoff positions of the atoms are listed in Table S1 in the Supporting Information, while the illustration of the structure is given in Figure 3a. There is no indication for peak splitting as it can be expected in case In and Sn are ordered on two different Wyckoff positions as in the distorted spinel structure with space group Imma.

\begin{figure}[h]
    \centering
    \includegraphics[width=\columnwidth]{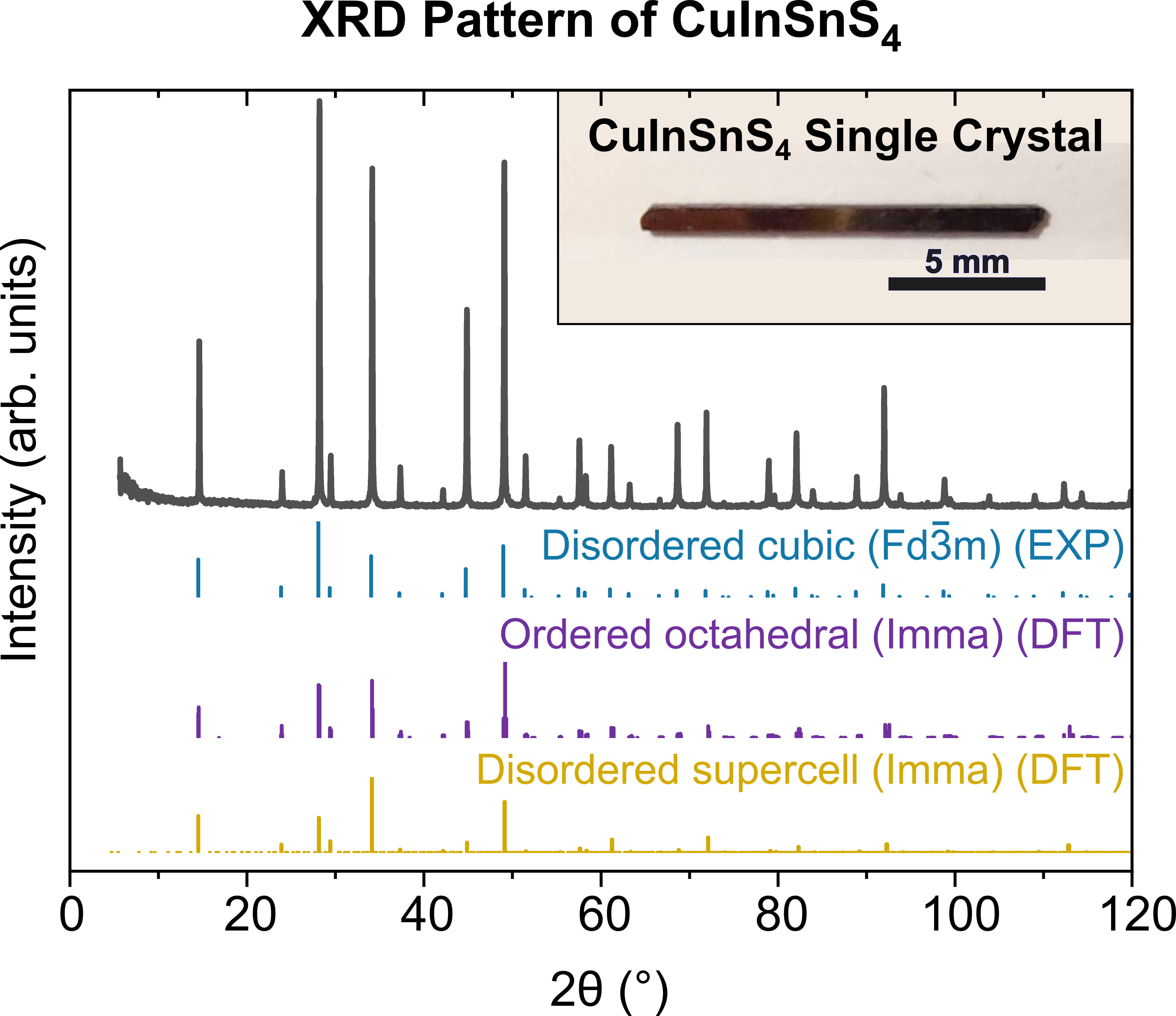}
    \caption{ Synchrotron X-ray diffraction patterns of CuInSnS$_4$ obtained from crushed single crystals and reference peak positions for different structural models. The experimental diffraction pattern (black dots) is well described by an average cubic spinel structure ($Fd\bar{3}m$). Vertical marks indicate Bragg peak positions for an ordered octahedral Imma structure (purple), a disordered Imma-derived supercell (orange), and a disordered cubic $Fd\bar{3}m$ structure (blue), highlighting the limited sensitivity of conventional XRD to In/Sn ordering. The inset shows an optical image of the CuInSnS$_4$ single crystal used in this study.}
    \label{crystal}
\end{figure}

\begin{figure*}[!t]
    \centering
    \includegraphics[width=\textwidth]{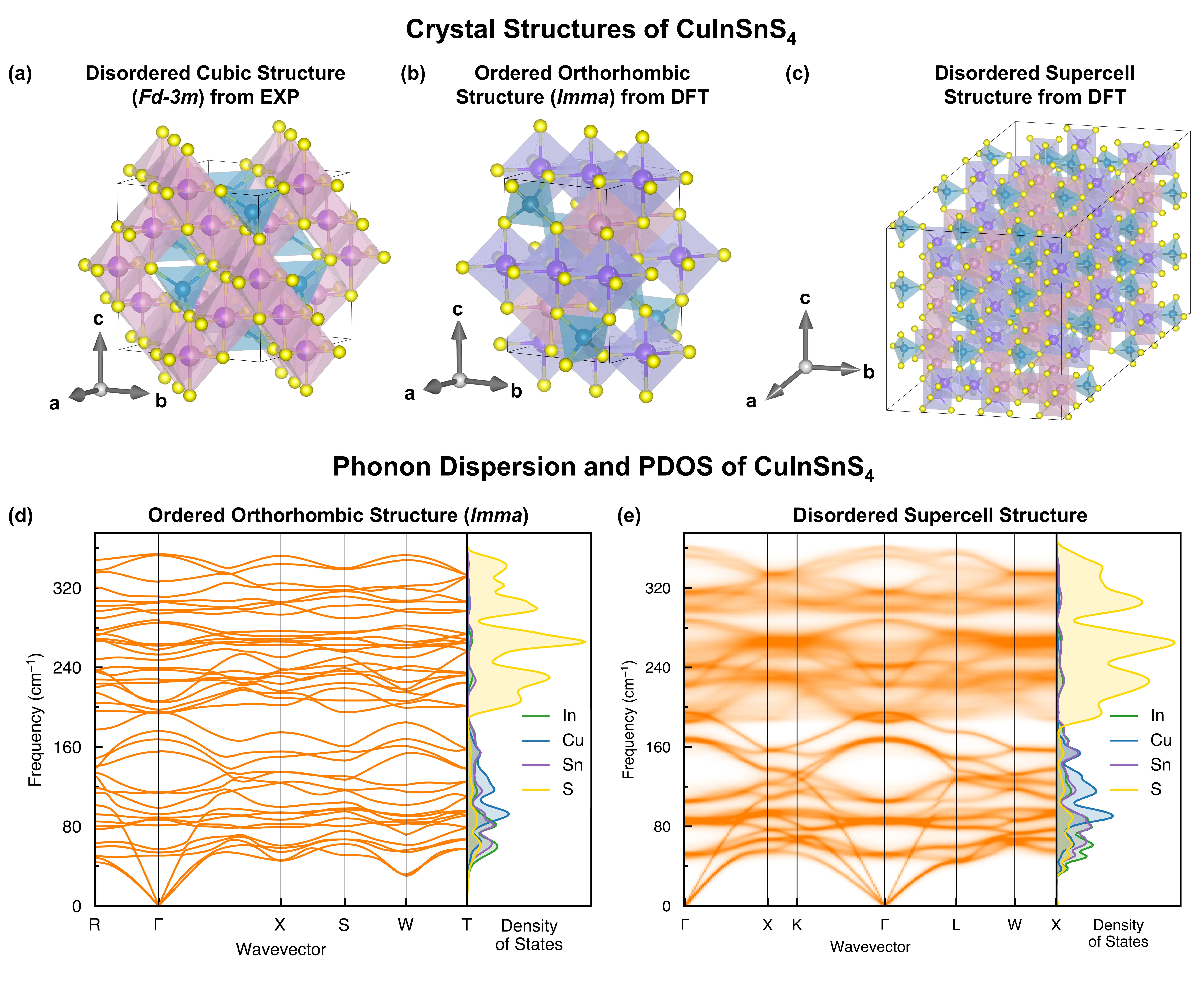}
    \caption{\textbf{Structural models and lattice dynamics of CuInSnS$_4$.} 
(a) Experimentally refined average cubic spinel structure (Fd$\bar{3}$m) obtained from XRD. 
(b) DFT-optimized ordered octahedral structure with Imma symmetry. 
(c) Disordered supercell derived from the Imma structure, with In and Sn quasi-randomly distributed over octahedral sites while preserving the cubic structure on average. 
(d,e) Calculated phonon dispersions and atom-projected phonon density of states (PDOS) for the ordered Imma structure and the disordered supercell, respectively.}
    \label{fig:rphonondisp}
\end{figure*}

To investigate the structural consequences of In/Sn ordering/disordering, we have performed lattice-dynamical calculations for multiple structural models. First, we consider cation-ordered octahedral structure with Imma symmetry, in which In and Sn occupy distinct octahedral sublattices. As illustrated in Figure 3a, this ordering leads to alternating InS$_6$ and SnS$_6$ octahedra and lowers the symmetry relative to the cubic spinel. In the ordered Imma structure, the In- and Sn-centered octahedra adopt direction-dependent metal--sulfur bond lengths, lowering the symmetry from cubic to orthorhombic and resulting in lattice parameters $a  \approx 7.371$~\AA\, $b \approx 7.415$~\AA\ and $c \approx 10.439$~\AA. The Wyckoff positions of the atoms are listed in Table S2 in the Supporting Information.

To emulate the experimentally observed cubic structure while incorporating intrinsic cation disorder, a 378-atom $3\times3\times3$ supercell was constructed based on the ordered Imma structure, with In and Sn (quasi-)randomly distributed over octahedral sites to recover an average Fd$\bar{3}$m symmetry while introducing local structural variations. As shown in Figure 3b, this disordered configuration preserves the cubic metric on average, with an effective cubic lattice parameter of $a \approx 10.451$~\AA\ calculated using the PBEsol semi-local DFT functional, very slightly underestimating the experimental value ($a = 10.495 \pm 0.002$~\AA). At the same time, the random In/Sn distribution breaks local symmetry, such that individual InS$_6$ and SnS$_6$ octahedra are no longer symmetry-equivalent and exhibit locally distorted bonding environments that are absent in the ordered Imma model.

We find a relatively modest impact of cation disorder on the lattice dynamics of CuInSnS$_4$, as directly shown in both the radial distribution functions (RDF) and the phonon dispersions. Comparison of the ordered Imma and disordered supercell RDFs shows that local bonding environments are remarkably similar in the two structures (Figure S1 in the Supporting Information). In both cases, the first-neighbor Cu--S distances are narrowly distributed around approximately 2.3--2.4~\AA, while In--S and Sn--S bonds span a slightly broader but still well-defined range of roughly 2.4--2.6~\AA. These nearly identical first-shell bond lengths indicate that short-range coordination is preserved upon disordering, despite the loss of long-range cation periodicity. Differences between the ordered and disordered structures emerge primarily at intermediate distances, where the disordered structure exhibits broadened and partially merged peaks in the 4.2--4.6~\AA\ range, reflecting locally disordered octahedral connectivity rather than a fundamental change in bonding topology.

This preservation of local bonding is observed in the phonon density of states (PDOS) and dispersions. As shown in Figure~3 d,e, the PDOS of the disordered structure closely resembles that of the ordered Imma structure, with only moderate broadening of individual features. In both cases, a clear mass-dependent hierarchy is retained, with low-frequency modes below approximately 120~cm$^{-1}$ dominated by In and Sn motion, intermediate modes between about 120 and 220~cm$^{-1}$ involving primarily Cu vibrations, and high-frequency optical modes above roughly 220~cm$^{-1}$ dominated by sulfur motion. Disorder primarily manifests as a mild broadening of phonon branches and DOS features, most noticeably in the high-frequency sulfur-dominated optical modes, consistent with the greater sensitivity of short-wavelength vibrations to local structural variations.

The relatively weak phonon broadening induced by disorder in CuInSnS$_4$ contrasts with the behavior observed in several other cation-disordered multinary semiconductors, such as AgBiS$_2$,\cite{niedziela2020controlling} AgSbTe$_2$,\cite{ye2008first} and related I–V–VI$_2$ chalcogenides, as well as disordered kesterites like Cu$_2$ZnSn(S,Se)$_4$,\cite{schorr2019point} where sublattice disorder is commonly associated with strong phonon scattering, pronounced linewidth broadening, and reduced vibrational coherence. In CuInSnS$_4$, this muted response can be traced to the close similarity between In and Sn, which are adjacent in the periodic table and have nearly identical atomic masses and ionic radii, along with similar valence electronic configurations. As a result, metal--sulfur bond lengths and coordination environments remain nearly indistinguishable upon cation mixing, as shown by the RDFs. This behavior is consistent with the low energetic cost of disordering, with the calculated enthalpy difference between ordered and disordered configurations on the order of a few meV per atom ($\sim\SI{7.5}{meV/atom}$), corresponding to an estimated thermodynamic order--disorder transition temperature of a few hundred kelvin.

While disorder has a limited impact on local bonding strength and phonon energies, it nevertheless has important consequences for vibrational symmetry. In the ordered Imma structure, lattice vibrations can be classified according to well-defined irreducible representations, yielding A$_g$, B$_{1g}$, B$_{2g}$, and B$_{3g}$ modes that are symmetry-distinct and non-degenerate. When cation disorder restores an average cubic metric consistent with Fd$\bar{3}$m symmetry, these symmetry distinctions are no longer strictly preserved. Instead, phonon modes derived from the ordered structure mix and evolve toward the more degenerate vibrational manifolds expected for the cubic lattice, with local symmetry breaking further relaxing polarization and activity selection rules. As a result, the experimentally observed vibrational response reflects an effective symmetry that lies between the ideal ordered and fully cubic limits. In the next section, we examine in detail these disorder-induced changes in phonon symmetry and selection rules using multi-wavelength excitation and polarization-dependent Raman spectroscopy.

\subsection{Vibrational Properties}

We now turn to the vibrational properties of CuInSnS$_4$ by considering the symmetry-allowed Raman-active phonon modes for the two relevant structural models. For the average cubic spinel structure with space group Fd$\bar{3}$m and point group O$_h$ (m$\bar{3}$m), group-theoretical analysis predicts that the Raman-active zone-center vibrations decompose as
\begin{equation}
\Gamma_{\mathrm{Raman}}^{\mathrm{Fd\bar{3}m}} = A_{1g} + E_g + 3T_{2g},
\end{equation}
corresponding to a total of five Raman-active phonon modes. Here, the labels $A$, $E$, and $T$ denote non-degenerate, doubly degenerate, and triply degenerate vibrational modes, respectively. The $A$ and $B$ labels denote symmetric and antisymmetric vibrational modes, respectively. The subscript $g$ indicates even parity under inversion, which is a requirement for Raman activity in centrosymmetric crystals. The numerical subscripts (1 and 2) distinguish modes of the same degeneracy but different symmetry with respect to the crystal’s rotational operations, and therefore correspond to different Raman tensor forms.

For a fully ordered orthorhombic structure with space group Imma and point group D$_{2h}$ (mmm), a larger number of Raman-active modes is expected due to the lifting of the cubic degeneracies. In this case, the Raman-active phonons transform according to
\begin{equation}
\Gamma_{\mathrm{Raman}}^{\mathrm{Imma}} = 5A_g + 2B_{1g} + 4B_{2g} + 4B_{3g},
\end{equation}
yielding a total of fifteen Raman-active modes. In the orthorhombic notation, all Raman-active modes belong to one-dimensional irreducible representations, with the subscripts 1, 2, and 3 distinguishing the different symmetry characters associated with rotations about the principal crystallographic axes. The complete mechanical representations for both structural models, including infrared-active and silent modes, are provided in the Supporting Information.

In the following sections, we investigate how CuInSnS$_4$ behaves from a vibrational point of view by employing multi-wavelength Raman excitation to determine the number of experimentally observed phonon modes, followed by polarization-dependent Raman spectroscopy to identify their symmetries.

\subsection{Multiwavelength Excitation Raman Measurements of CuInSnS$_4$}

Figure~4a schematically illustrates the excitation-wavelength-dependent penetration depth in CuInSnS$_4$, calculated from the experimentally reported absorption spectra.\cite{ramasamy2018nanocrystals} Because the absorption coefficient increases strongly toward higher photon energies, shorter-wavelength excitation results in shallower penetration depths and enhanced sensitivity to near-surface regions, whereas longer-wavelength excitation probes a larger effective volume of the bulk crystal. Figure~4b shows the corresponding multi-wavelength Raman spectra acquired using excitation wavelengths of 325, 405, 488, 532, and 785~nm, revealing both first-order and higher-order phonon features.

The pronounced dependence of Raman intensities on excitation wavelength reflects resonance effects governed by the electronic structure of CuInSnS$_4$. The DFT-calculated band structure and the absorption spectrum of the ordered Imma phase are shown in Figure~S2 in the Supporting Information. Based on these, we can observe energy levels at approximately 1.67, 2.68, 2.92, and 3.95~eV above the valence band at the Gamma point. These energies overlap selectively with the photon energies of the laser excitations used here, enabling resonance enhancement of specific phonon modes when the excitation energy is close to the energy band values.

\begin{figure}[H]
    \centering
    \includegraphics[scale=0.9]{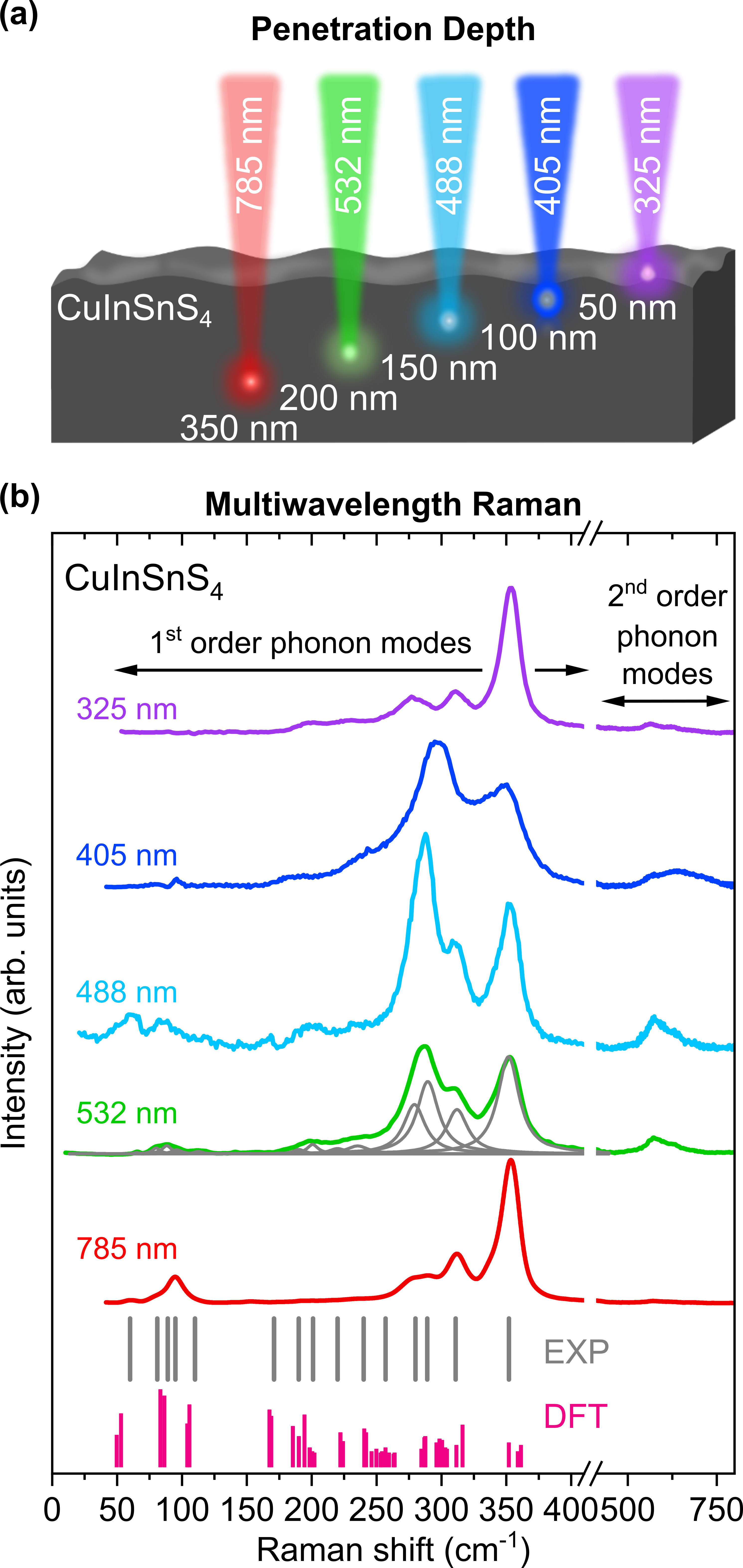}
    \caption{\textbf{Multi-wavelength excitation-dependent Raman measurements of CuInSnS$_4$.}
    (a) Schematic illustration of the excitation-wavelength-dependent penetration depth in CuInSnS$_4$, calculated from the absorption spectrum.
    (b) Raman spectra of CuInSnS$_4$ measured using 785, 532, 488, 405, and 325~nm excitation wavelengths, showing the evolution of first-order and higher-order phonon features with excitation energy. Representative deconvolution of the spectra is indicated in the Raman spectrum measured with 532~nm excitation. Vertical lines indicate the experimentally extracted peak positions (EXP) and the DFT-calculated phonon mode frequencies (DFT), as indicated in the bottom panel. The full list of DFT-calculated phonon modes, including their frequency and relative intensity, is given in Table~S3 in the Supporting Information.}
    \label{phonondisp}
\end{figure}

Under 785~nm excitation (1.58~eV), which lies close to the lowest-energy interband transition, the low-frequency modes in the $\sim$60--120~cm$^{-1}$ range exhibit comparatively higher intensity than under the other excitation wavelengths. As the excitation energy is increased to 532~nm (2.33~eV) and 488~nm (2.54~eV), approaching the higher-lying interband transition near 2.7~eV, the Raman response becomes dominated by first-order optical phonons in the $\sim$250--380~cm$^{-1}$ range, with the emergence of a clear multi-peak structure. In particular, modes near $\sim$290~cm$^{-1}$ and in the $\sim$340--360~cm$^{-1}$ region show a pronounced enhancement in intensity.

At even higher excitation energies, for 405~nm (3.06~eV) and 325~nm (3.82~eV), the spectra evolve toward a broader and more strongly enhanced response across the mid- and high-frequency range. The persistence and further enhancement of the $\sim$350~cm$^{-1}$ band at these excitation energies is consistent with resonance involving higher-lying electronic transitions with stronger sulfur $p$-orbital character -- consistent with the DFT band structure and density of states (Figure~S2). Overall, the excitation-energy-dependent evolution of the Raman spectra reveals mode-specific electron--phonon coupling. Low-frequency modes involving primarily In and Sn motion are selectively enhanced near the band edge, consistent with coupling to electronic states formed from hybridized In/Sn $s$ and S $p$ orbitals. Cu-related modes are most prominent at intermediate excitation energies. At higher excitation energies, sulfur-dominated optical modes become increasingly resonant as transitions with stronger S $p$ character are reached.

To identify the individual vibrational modes and distinguish between first-order and multi-phonon contributions, a detailed deconvolution of the Raman spectra was performed. The spectra shown in Figure~4b were fitted simultaneously for all excitation wavelengths using Lorentzian line shapes. Peak positions and linewidths for a given phonon mode were constrained to be identical across all spectra, while the intensities were treated as free parameters to capture excitation-dependent resonance effects.\cite{nielsen2025bazrs3, blaga2024unveiling} Initial peak positions were guided by phonon frequencies from first-principles calculations, and additional broader features were introduced to account for second-order phonon processes. As linewidths are primarily governed by phonon lifetimes and crystal quality, modes of the same phonon order were assumed to have comparable widths, while multi-phonon features were allowed to be broader.

Based on this analysis, fifteen first-order Raman-active modes were identified. Table 1 summarizes the experimentally obtained Raman mode frequencies and their comparison with the corresponding DFT-calculated values. The experimentally determined peak positions show good agreement with the DFT-calculated phonon modes of the disordered structure, as indicated by the comparison shown in the bottom panel of Figure~4b. 
On average, the frequencies of the lower energy modes tend to be slightly under-estimated by the PBEsol semi-local DFT functional, likely due to under-binding of semi-valence cation $d^{10}$ states, which softens the cation-anion bonds and is likely related to the (slightly) under-estimated lattice constant.\cite{nicolson_cu2sise3_2023,kavanagh_hidden_2021}
The higher-frequency features in Figure~4b are attributed to second-order Raman modes arising from overtones and combination modes of the identified first-order phonons, consistent with their broader linewidths and excitation-dependent intensities. We next employ polarization-dependent Raman spectroscopy to determine the symmetry of the observed Raman-active modes.

\begin{table}[ht]
\centering
\caption{Frequency (in cm$^{-1}$) of peaks from Lorentzian deconvolution of CuInSnS$_4$ Raman spectra and assigned mode symmetry from polarization-dependent Raman measurements compared with theoretical DFT calculations of the disordered structure and experimental conditions under which the modes are best resolved.}
\label{tab:dft_exp_mode_match}
\renewcommand{\arraystretch}{1.15}
\begin{tabular}{c c c c}
\hline
\textbf{$\nu_{\mathrm{EXP}}$} & \textbf{Symmetry} & \textbf{$\nu_{\mathrm{DFT}}$} & \textbf{Excitation wavelength} \\
\textbf{(cm$^{-1}$)} &  & \textbf{(cm$^{-1}$)} & \textbf{(nm)} \\
\hline

60 & $T_{2g}-like$ & 53  & 785 \\
81 & $T_{2g}-like$ & 83  & 785 \\
89 & $T_{2g}-like$ & 86  & 785 \\
95 & $T_{2g}-like$ & 104 & 488, 532, 785 \\
110 & $T_{2g}-like$ & 106 & 488, 532 \\
171 & $T_{2g}-like$ & 167 & 325, 405, 488 \\
190 & $T_{2g}-like$ & 190 & 325, 405, 488 \\
201 & $T_{2g}-like$ & 200 & 325, 405 \\
220 & $T_{2g}-like$ & 222 & 325, 405 \\
240 & $T_{2g}-like$ & 240 & 405, 488, 532 \\
257 & $T_{2g}-like$ & 257 & 325, 405 \\
280 & $T_{2g}-like$ & 285 & 325, 405, 488, 532, 785 \\
289 & $T_{2g}-like$ & 287 & 325, 405, 488, 532, 785 \\
311 & $T_{2g}-like$ & 312 & 325, 405, 488 \\
353 & $T_{2g}-like$ & 352 & 325 \\
\hline
\end{tabular}
\end{table}

\subsection{Angle-Resolved Raman Polarization Measurements of CuInSnS$_4$}

To determine the symmetry of the Raman-active phonon modes, we employ polarization-dependent Raman spectroscopy and analyze the angular dependence of the scattered intensity within the framework of Raman tensors. Under non-resonant conditions, the Raman response of each vibrational mode is governed by its corresponding Raman tensor, which carries the symmetry of the phonon and dictates the polarization selection rules.

For the Raman-active modes of both the cubic spinel ($Fd\bar{3}m$) and orthorhombic Imma structures,\cite{gallego2019automatic} the Raman tensors are fully symmetric and take diagonal or single-component forms depending on the symmetry of the mode. For cubic $Fd\bar{3}m$, the Raman tensors are given by
\begin{equation}
\begin{aligned}
\mathcal{R}(A_{1g}) &=
\begin{pmatrix}
a & 0 & 0\\
0 & a & 0\\
0 & 0 & a
\end{pmatrix}, \quad
\mathcal{R}(E_g) =
\begin{pmatrix}
b & 0 & 0\\
0 & c & 0\\
0 & 0 & d
\end{pmatrix}, \\[6pt]
\mathcal{R}(T_{2g}) &=
\begin{pmatrix}
0 & e & e\\
e & 0 & e\\
e & e & 0
\end{pmatrix}.
\end{aligned}
\end{equation}

where $a$, $b$, $c$, $d$, and $e$ are the Raman tensor coefficients, while for orthorhombic Imma the Raman tensors are:

\begin{equation}
\begin{aligned}
\mathcal{R}(A_g) &=
\begin{pmatrix}
a & 0 & 0 \\
0 & b & 0 \\
0 & 0 & c
\end{pmatrix}, \quad
\mathcal{R}(B_{1g}) =
\begin{pmatrix}
0 & d & 0 \\
d & 0 & 0 \\
0 & 0 & 0
\end{pmatrix}, \\[6pt]
\mathcal{R}(B_{2g}) &=
\begin{pmatrix}
0 & 0 & e \\
0 & 0 & 0 \\
e & 0 & 0
\end{pmatrix}, \quad
\mathcal{R}(B_{3g}) =
\begin{pmatrix}
0 & 0 & 0 \\
0 & 0 & f \\
0 & f & 0
\end{pmatrix}.
\end{aligned}
\end{equation}

where $a$, $b$, $c$, $d$, $e$, and $f$ are the Raman tensor coefficients. In contrast to the cubic case, all degeneracies are lifted in the Imma structure, and each Raman-active mode exhibits a unique polarization dependence.

The Raman scattering intensity \cite{cardona2006light,flor2022raman} for a given phonon mode is proportional to
\begin{equation}
I_s \propto C(\omega_p)\, \left| \hat{e}_i \cdot \mathcal{R} \cdot \hat{e}_s \right|^2,
\end{equation}
where $\hat{e}_i$ and $\hat{e}_s$ are the polarization unit vectors of the incident and scattered light, respectively, and $\mathcal{R}$ is the Raman tensor of the corresponding mode. The prefactor $C(\omega_p)$ accounts for the frequency dependence of the Raman process and is given by
\begin{equation}
C(\omega_p) = \frac{\omega_i (\omega_i - \omega_p)^3}{\omega_p \left[1-\exp(-\hbar \omega_p / k_B T)\right]},
\end{equation}
where $\omega_i$ is the excitation frequency, $\omega_p$ is the phonon frequency, $\hbar$ is the reduced Planck constant, $k_B$ is the Boltzmann constant, and $T$ is the temperature.

Polarization-dependent Raman measurements were performed on a CuInSnS$_4$ single crystal with the (010) plane as the basal surface, as shown in Figure 5. The polarization vectors of the incident and scattered light were therefore defined within the $ac$ plane as

\begin{equation}
\begin{aligned}
\hat{\mathbf{e}}_i &=
\begin{pmatrix}
\sin\theta \\
0 \\
\cos\theta
\end{pmatrix}, \quad
\hat{\mathbf{e}}_s^{\parallel} =
\begin{pmatrix}
\sin\theta \\
0 \\
\cos\theta
\end{pmatrix}, \quad
\hat{\mathbf{e}}_s^{\perp} =
\begin{pmatrix}
\cos\theta \\
0 \\
-\sin\theta
\end{pmatrix}.
\end{aligned}
\end{equation}

where $\theta$ denotes the polarization angle with respect to the [100] crystallographic direction, and the superscripts $\parallel$ and $\perp$ correspond to parallel and cross-polarized detection geometries, respectively.

Combining Eqs.~(3-7) yields the angular dependence of the Raman intensities for the different phonon symmetries, as summarized in Table~2 for measurements performed on the (010) basal plane. For the cubic Fd$\bar{3}$m structure, the fully symmetric $A_{1g}$ mode exhibits a polarization-independent intensity in the parallel configuration, due to its isotropic Raman tensor. The $E_g$ modes display a characteristic angular dependence governed by $\cos^2\theta$ and $\sin^2\theta$ terms, reflecting the anisotropic diagonal elements of the $E_g$ Raman tensor, while the $T_{2g}$ modes are active in both parallel and crossed polarization and show an angular dependence proportional to $\sin^2(2\theta)$, arising from the off-diagonal tensor components. Because of the cubic symmetry, the qualitative angular dependence of these Raman modes is invariant with respect to the choice of crystallographic basal plane, and equivalent polarization behavior is expected for measurements performed on the (100), (010), or (001) planes.

In the case of the orthorhombic Imma structure, the lowering of symmetry lifts the cubic degeneracies and leads to distinct $A_g$ and $B_{ig}$ ($i = 1, 2, 3$) Raman-active modes with direction-dependent Raman tensors. As a result, the polarization response depends explicitly on the choice of the crystallographic plane. In the (010) scattering geometry considered here, in the parallel polarization configuration only $A_g$ and $B_{2g}$ modes are symmetry allowed, with the $B_{2g}$ modes showing a $\sin^2(2\theta)$ angular dependence identical to that of the cubic $T_{2g}$ modes. For measurements on the (100) or (001) planes, different subsets of $B_{ig}$ modes would become active, and the corresponding angular dependencies would change accordingly. This plane-dependent behavior is a direct consequence of the reduced orthorhombic symmetry and contrasts with the plane-invariant response expected for the cubic structure.

In the perpendicular (crossed) polarization configuration, the selection rules are modified relative to the parallel case. For the cubic Fd$\bar{3}$m structure, the fully symmetric $A_{1g}$ mode becomes symmetry forbidden, while the $E_g$ modes remain allowed and exhibit an angular dependence governed by mixed $\sin^2\theta\cos^2\theta$ terms, reflecting the anisotropy of the diagonal Raman tensor elements. The $T_{2g}$ modes remain strongly allowed in crossed polarization and display a pronounced $\cos^2(2\theta)$ angular dependence, consistent with their off-diagonal Raman tensor components.

For the orthorhombic Imma structure, in the (010) scattering geometry considered here, the $A_g$ modes remain allowed and exhibit an angular dependence governed by mixed $\sin^2\theta\cos^2\theta$ terms, just like the $E_g$ modes for the cubic Fd$\bar{3}$m structure. The $B_{2g}$ modes also remain allowed and exhibit the same $\cos^2(2\theta)$ angular dependence for crossed polarization as the cubic $T_{2g}$ modes. Modes of $B_{1g}$ and $B_{3g}$ symmetry are forbidden in this geometry due to the absence of polarization components along the $b$ axis. As in the parallel configuration, the identical angular dependencies of the $B_{2g}$ and $T_{2g}$ modes imply that crossed-polarization Raman measurements on a single crystallographic plane cannot distinguish between these two symmetries.

\begin{table}[h]
\centering
\caption{Angular dependence of Raman mode intensity for CuInSnS$_4$ on the (010) plane for the cubic Fd$\bar{3}$m and orthorhombic Imma structures.}
\renewcommand{\arraystretch}{1.15}
\begin{tabular}{lcc}
\toprule
Mode (structure) & Parallel & Perpendicular \\
\midrule
$A_{1g}$ (Fd$\bar{3}$m) & $a^2$ & $0$ \\
$E_g$ (Fd$\bar{3}$m) & $(b\cos^2\theta+d\sin^2\theta)^2$ & $(b-d)^2\sin^2\theta\cos^2\theta$ \\
$T_{2g}$ (Fd$\bar{3}$m) & $e^2\sin^2(2\theta)$ & $e^2\cos^2(2\theta)$ \\
\midrule
$A_g$ (Imma) & $(a\sin^2\theta+c\cos^2\theta)^2$ & $(a-c)^2\sin^2\theta\cos^2\theta$ \\
$B_{1g}$ (Imma) & $0$ & $0$ \\
$B_{2g}$ (Imma) & $e^2\sin^2(2\theta)$ & $e^2\cos^2(2\theta)$ \\
$B_{3g}$ (Imma) & $0$ & $0$ \\
\bottomrule
\end{tabular}
\label{tab:pol_singlecol}
\end{table}

Figure~6 shows the polarization-dependent Raman spectra of the CuInSnS$_4$ single crystal measured at 290~K using two excitation wavelengths (785 and 532~nm). The measurements were performed on the (010) basal plane in both parallel ($\hat{e}_i \parallel \hat{e}_s$) and crossed ($\hat{e}_i \perp \hat{e}_s$) polarization configurations, as shown in panels (a,b) and (d,e), respectively. Identical polarization-dependent spectra were also obtained on the (100) plane, with no observable changes in peak positions, relative intensities, or angular dependencies, indicating that the polarization response is insensitive to the choice of basal plane.

Across both excitation wavelengths, all experimentally resolved Raman modes show the same dominant angular modulation, consistent with a single effective symmetry. This behavior is presented in Figure~6(c,f), where the intensity of a representative Raman mode at 353~cm$^{-1}$ follows a $\sin^2(2\theta)$ dependence in the parallel configuration and a $\cos^2(2\theta)$ dependence in the crossed configuration. This angular response is governed by off-diagonal Raman tensor elements and is formally identical to the behavior expected for cubic $T_{2g}$ modes, as well as for orthorhombic $B_{ig}$ modes of the Imma structure when probed in this configuration. The persistence of the same angular dependence for all modes and excitation wavelengths demonstrates that the polarization-dependent response is isotropic in symmetry, even though resonance effects redistribute the spectral weight.

Interestingly, the number of experimentally observed first-order Raman peaks is fifteen, exceeding the five Raman-active modes allowed by cubic Fd$\bar{3}$m symmetry and matching the number of Raman-active modes predicted for the ordered Imma structure. This indicates that, from the number of modes perspective, the vibrational spectrum of CuInSnS$_4$ reflects a reduced local symmetry in which phonon degeneracies are lifted. At the same time, the polarization-dependent measurements reveal that all observed modes share the same off-diagonal ($T_{2g}/B_{2g}$-like) angular dependence, implying that the vibrational response is effectively isotropic in symmetry when averaged over the probed volume.

From a phonon perspective, this means that CuInSnS$_4$ does not behave as a uniformly orthorhombic crystal with symmetry-pure modes, nor as an ideal cubic spinel. Instead, local symmetry breaking associated with intrinsic In/Sn disorder splits and activates additional phonon modes, increasing the number of observable vibrations, while disorder-induced mode mixing and spatial averaging cause the Raman tensors of these modes to be dominated by off-diagonal components. As a result, phonons originating from different irreducible representations acquire similar effective Raman tensor forms and therefore exhibit indistinguishable polarization behavior.

\begin{figure}[H]
    \centering
    \includegraphics[scale=0.8]{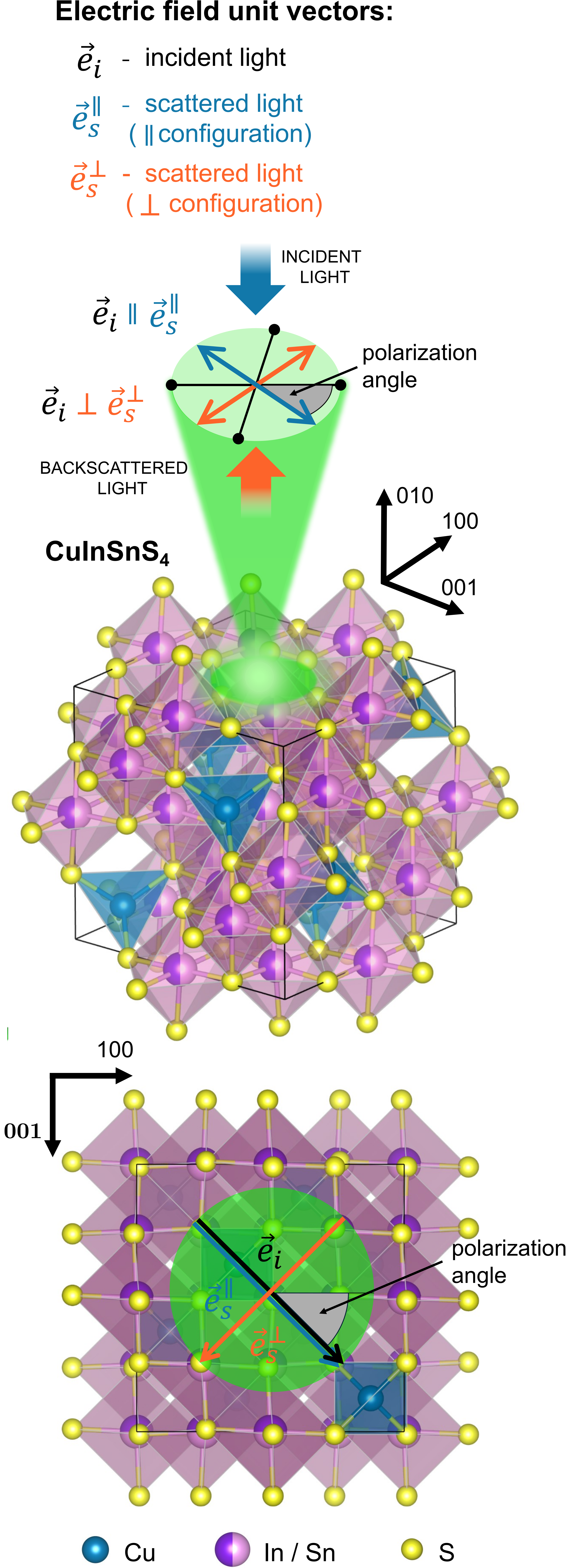}
    \caption{Schematic illustration of the polarized Raman scattering geometry used in this work. 
Polarization-dependent Raman measurements were performed in backscattering configuration on the (010) basal plane, with the incident laser beam propagating normal to the surface. The polarization direction of the incident light $\hat{e}_i$ was rotated with respect to the crystallographic axes, while the scattered light was analyzed in parallel ($\hat{e}_i \parallel \hat{e}_s$) and crossed ($\hat{e}_i \perp \hat{e}_s$) configurations, as indicated by the blue and orange arrows, respectively. The crystallographic orientations are marked, and the atomic structure of CuInSnS$_4$ is shown for reference, with Cu, In/Sn, and S atoms indicated by different colors.}

    \label{crystalpol}
\end{figure}

This behavior does not imply that $A_g$ modes are converted into $B_{2g}$ modes, nor that the agreement with the number of Imma phonon modes is coincidental. Rather, it reflects the fact that in a disordered lattice the vibrational eigenmodes are no longer symmetry-pure and cannot be uniquely classified by the irreducible representations of an ideal crystal. The fifteen experimentally observed modes therefore represent disorder-split and mixed phonons whose frequencies retain sensitivity to local symmetry lowering, while their polarization response reflects an averaged, cubic-like vibrational anisotropy. In this sense, CuInSnS$_4$ exhibits locally anisotropic but macroscopically isotropic vibrational behavior.

\FloatBarrier
\begin{figure*}[t]
    \centering
    \includegraphics[width=\textwidth]{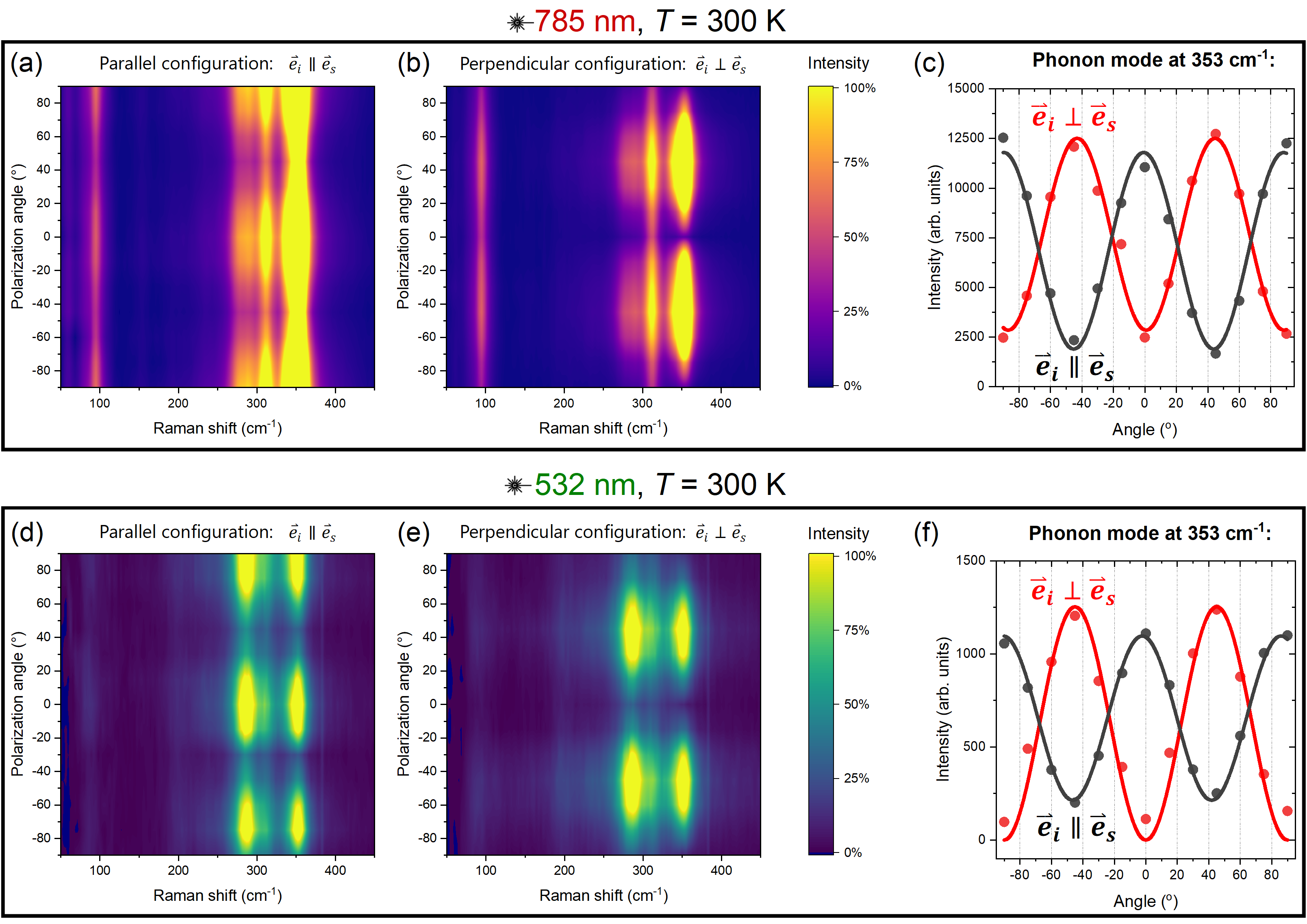}
   \caption{Polarization-dependent Raman response of CuInSnS$_4$ measured at 300 K with 785 nm and 532 nm excitation in parallel ($\mathbf{\bar e}_i \parallel \mathbf{\bar e}_s$) and perpendicular ($\mathbf{\bar e}_i \perp \mathbf{\bar e}_s$) polarization configurations. (a,b) Raman intensity maps as a function of polarization angle for 785 nm excitation in the parallel and perpendicular configurations, respectively. (c) Angular dependence of the integrated Raman area  of the 353 cm$^{-1}$ mode obtained from deconvolution the spectra shown in (a) and (b), plotted for the parallel (black) and perpendicular (red) configurations. (d,e) Raman intensity maps as a function of polarization angle for 532 nm excitation in the parallel and perpendicular configurations, respectively. (f) Angular dependence of the integrated Raman area  of the 353 cm$^{-1}$ mode obtained from deconvolution the spectra shown in (d) and (e), plotted for the parallel (black) and perpendicular (red) configurations. Solid lines are fits using the expressions given in Table~2.}
    \label{fig:rpol}
\end{figure*}

The coexistence of an Imma-like Raman mode number with a largely uniform ($T_{2g}/B_{2g}$-like) polarization response is consistent with prior observations in disordered multinary and spinel-derived materials, where local symmetry breaking lifts degeneracies and activates additional modes while disorder-induced mode mixing and averaging shadow the mode-resolved selection rules. \cite{Ivanov2010,Ursaki2002,Lazzeri2006} Analogous limitations of symmetry-pure assignments from polarized Raman data have also been discussed for strongly disordered multinary chalcogenides such as kesterites, where local cation disorder affects both mode visibility and polarization selection rules.\cite{Djemour2013,Guc2016} More generally, the notion that Raman spectra can reflect an ``effective'' symmetry in the presence of local disorder has been emphasized in halide perovskites, where disorder and anharmonic fluctuations relax selection rules and complicate a strict factor-group interpretation.\cite{Menahem2023}

\subsection{Photoluminescence Spectroscopy}

We now turn to the optoelectronic properties and carrier dynamics of CuInSnS$_4$ as measured using steady-state and time-resolved PL spectroscopy.

Figure~7 summarizes the steady-state and time-resolved photoluminescence (PL and TRPL) response of the CuInSnS$_4$ single crystal. The PL spectrum is dominated by a broad, asymmetric emission band that can be deconvoluted into two Gaussian components, corresponding to a higher-energy and a lower-energy transitions. As discussed below, these two features exhibit distinct excitation-density, temperature, polarization, and temporal dynamics, indicating different microscopic origins.

The higher-energy emission peak is centered at 1.58 eV, closely matching the fundamental band gap at 1.65 eV of the ordered Imma structure from hybrid DFT calculations. The neglect of In/Sn disorder and thermal effects likely contribute to the slightly larger band gap from the ordered phase DFT calculation. This energy also agrees well with reported optical band gaps extracted from absorption measurements on CuInSnS$_4$ in the literature.\cite{chai2023metal,ELRADAF2023414867} This indicates that the higher-energy PL feature is probably associated with band-edge recombination. The lower-energy emission around 1.48 eV appears as a red-shifted shoulder and is indicative of recombination involving disorder-related states. In order to determine the exact origin of these transitions we use the power- and temperature dependent PL measurements.

\subsubsection{Power-dependent photoluminescence}

Power-dependent PL measurements were performed by varying the excitation density in the range from 10 to 170~mW\,cm$^{-2}$ at a constant temperature of 80~K. With increasing excitation density, the overall PL intensity increases while the spectral shape remains largely unchanged, indicating the absence of pronounced state filling, spectral saturation, or strong heating effects within the explored power range.

We determine the microscopic origin of the two emission channels by analyzing the excitation-density dependence of their integrated PL intensities using the empirical power-law relation
\begin{equation}
I \propto P^{k},
\end{equation}
where $I$ is the integrated PL intensity, $P$ is the excitation density, and $k$ reflects how the radiative recombination rate scales with the photogenerated carrier density.\cite{nielsen2025bazrs3,nielsen2026spectroscopic} In general, a $k \geq 1$ are associated with band-edge recombination processes, including band-to-band or excitonic emission, for which the radiative rate scales approximately linearly with carrier density. Sublinear behavior ($k < 1$) is commonly observed for defect-mediated recombination, where saturation of defect states, competing non-radiative pathways, or carrier trapping reduce the power dependence.

Fits to the experimental data yield a power-law exponent of $k_{\mathrm{BB}} = 1.20 \pm 0.11$ for the high-energy emission, which we label as the band-to-band (BB) transition. This value is consistent with predominantly band-edge recombination involving free or weakly bound excitons. In contrast, the low-energy contribution, labeled as the DX transition, exhibits a near-linear power dependence with an exponent of $k_{\mathrm{DX}} = 0.97 \pm 0.20$. Within experimental uncertainty, this behavior is compatible with several distinct recombination mechanisms, including free-to-bound or donor--acceptor-pair transitions that do not saturate within the investigated power range, recombination via shallow defect states, and emission from disorder-induced band-tail or weakly localized excitonic states that remain partially coupled to the band edge.

While the near-linear power dependence of the DX emission already rules out domination by deep, isolated defect levels, it does not by itself uniquely identify the underlying recombination mechanism. To further discriminate between the possible assignments, we  investigate the excitation-density dependence of the PL linewidth and peak position, as shown in Figure S3 in the Supporting Information.

The DX emission exhibits a pronounced and non-monotonic evolution of the linewidth with increasing excitation density. At low excitation densities, the DX linewidth increases, consistent with progressive filling of higher-energy localized states and an increased spread of emitting energies within a disorder-broadened potential landscape. At intermediate excitation densities, the linewidth tends to saturate, and at the highest excitation densities it decreases, indicating that increased carrier density partially screens the disorder potential and promotes partial delocalization, thereby reducing the energetic dispersion of the emitting states. Throughout this excitation range, the DX peak position remains largely invariant, demonstrating that increasing excitation density primarily modifies the energetic spread of the localized-state manifold rather than shifting the average recombination energy.

On the other hand, the BB emission shows only weak and smooth variations of the peak position and a monotonic broadening with increasing excitation density, consistent with band-edge or weakly excitonic recombination. The absence of systematic blueshifts further indicates that band filling remains negligible under the present excitation conditions.

By comparison, donor--acceptor-pair recombination would be expected to produce a clear excitation-induced blueshift of the emission peak due to preferential recombination of closer pairs at higher carrier densities, which is not observed. Likewise, emission from a single discrete defect level would typically show weak excitation dependence or early saturation of both linewidth and intensity. 

The combined linewidth and peak-position behavior therefore points that the DX emission originates from disorder-related band-tail or weakly localized excitonic states that remain partially coupled to the band edge.

\subsubsection{Temperature-dependent photoluminescence}

Additional insight into the nature of the two emission channels is obtained from temperature-dependent PL measurements, as shown in Figure S4 in the Supporting Information. The temperature evolution of the BB and DX peak positions is different, reflecting their distinct microscopic origins. The BB peak exhibits a monotonic redshift with increasing temperature, characteristic of band-edge emission governed by electron--phonon interactions and thermal bandgap shrinkage. This behavior is well described by the O’Donnell--Chen model \cite{ODonnellChen1991},
\begin{equation}
E_{\mathrm{BB}}(T) = E_{\mathrm{BB}}(0) - S\langle \hbar\omega \rangle
\left[
\coth\!\left(\frac{\langle \hbar\omega \rangle}{2k_{\mathrm{B}}T}\right) - 1
\right],
\label{eq:ODC}
\end{equation}
where $E_{\mathrm{BB}}(0)$ is the zero-temperature band-edge energy, $S$ is a dimensionless electron--phonon coupling constant, and $\langle \hbar\omega \rangle$ is an effective phonon energy. The fitting parameters extracted are summarized in Table~3. The coupling constant $S \approx 0.30$ is relatively modest compared to values typically reported for bulk chalcogenide semiconductors or III--V compounds \cite{isik2020temperature}, where $S$ values in the range of 2--3 have been observed, indicating stronger coupling to lattice vibrations in those systems. This comparatively low $S$ suggests that lattice vibrations play only a limited role in the band-edge transition in CuInSnS$_4$, pointing to weaker electron--phonon coupling than in many conventional semiconductors. More generally, in this regime the coupling is often dominated by low-energy optical phonon modes, which in the present case occur at approximately $80~\mathrm{cm}^{-1}$.

The DX peak position exhibits a weaker and non-monotonic temperature dependence that cannot be captured by a simple bandgap-shrinkage description. Instead, the DX emission is better described within a localized-state (band-tail) framework, in which structural disorder gives rise to a distribution of localized electronic states near the band edge. Within this model, the temperature dependence of the emission energy follows \cite{li2005origin}
\begin{equation}
E_{\mathrm{DX}}(T) = E_{\mathrm{DX}}(0) - \frac{\sigma^2}{k_{\mathrm{B}}T},
\label{eq:DX_model}
\end{equation}
where $E_{\mathrm{DX}}(0)$ is the characteristic low-temperature emission energy and $\sigma$ represents the energetic width of the localized-state distribution. This model describes how carriers redistribute thermally within localized states. At low temperature, recombination occurs predominantly from the lowest-energy localized states, while increasing temperature enables carriers to access higher-energy states prior to radiative recombination. As a result, significant changes in linewidth and intensity occur without a systematic shift of the average emission energy. The extracted values of $\sigma$ on the order of 10--20~meV (Table~3) indicate substantial energetic disorder, consistent with the local structural distortions and cation disorder also seen in  the vibrational analysis. This value is comparable in magnitude to the characteristic localization or tail energy scales (tens of meV) reported for a range of disordered semiconductors, indicating a moderate degree of electronic disorder rather than deep, isolated trapping. \cite{he2016exciton,roble2019impact,wright2017band}

\begin{table}[ht]
\centering
\caption{Fitting parameters for the temperature dependence of PL peak positions obtained using Eqs.~(\ref{eq:ODC}) and (\ref{eq:DX_model}).}
\label{tab:pl_peak_fits}
\renewcommand{\arraystretch}{1.2}
\begin{tabular}{lcc}
\toprule
Parameter & BB (band-to-band) & DX (disorder-related) \\
\midrule
$E_0$ (eV) & $ 1.56 \pm 0.03$ & $ 1.48 \pm 0.08$ \\
$S$ (--) & $ 0.30 \pm 0.08$ & -- \\
$\langle \hbar\omega \rangle$ (meV) &  $ 10 \pm 5$ & -- \\
$\sigma$ (meV) & -- & $ 15 \pm 5$ \\
\bottomrule
\end{tabular}
\end{table}

The temperature evolution of the PL peak widths as shown in Figure S4 in the Supporting Information provides an independent probe of carrier localization and electron–phonon coupling. The full width at half maximum (FWHM) of both BB and DX emissions increases with temperature and can be described by an exciton–phonon broadening model with an inhomogeneous baseline,\cite{nielsen2025parallel, nielsen2026spectroscopic}
\begin{equation}
\Gamma(T) = \Gamma_{\mathrm{inh}} + \gamma_{\mathrm{ac}} T
+ \frac{\Gamma_{\mathrm{LO}}}{\exp(E_{\mathrm{LO}}/k_BT)-1},
\label{eq:linewidth}
\end{equation}
where $\Gamma_{\mathrm{inh}}$ accounts for temperature-independent inhomogeneous broadening due to static disorder, $\gamma_{\mathrm{ac}}$ describes acoustic-phonon scattering, and the final term captures LO-phonon-assisted broadening.

The fitted parameters, summarized in Table~4, reveal stronger linewidth broadening for the DX emission compared to the BB emission, reflecting the enhanced sensitivity of the DX states to disorder and phonon scattering and consistent with carrier localization in a fluctuating potential landscape. For the BB transition, the linewidth evolution can be well described by a combination of inhomogeneous broadening and weak acoustic-phonon scattering.

For the DX transition the inclusion of an explicit LO-phonon coupling term does not improve the quality of the fit and leads to overparameterization, indicating that the linewidth broadening is dominated by disorder-related effects rather than by a well-defined LO-phonon interaction. In this case, the temperature dependence of the linewidth is governed primarily by inhomogeneous broadening and disorder-mediated carrier–phonon interactions, consistent with recombination from band-tail or weakly localized excitonic states.

\begin{table}[ht]
\centering
\caption{Fitting parameters for the temperature dependence of PL linewidths using an exciton–phonon broadening model with an inhomogeneous baseline.}
\label{tab:pl_width_fits}
\renewcommand{\arraystretch}{1.2}
\begin{tabular}{lcc}
\toprule
Parameter & BB (band-to-band) & DX (disorder-related) \\
\midrule
$\Gamma_{\mathrm{inh}}$ (meV) & $ 171 \pm 8$ & $ 210 \pm 6$ \\
$\gamma_{\mathrm{ac}}$ (meV/K) & $ 0.30 \pm 0.05$ & $ / $ \\
$\Gamma_{\mathrm{LO}}$ (meV) & $ / $ & $ 10 \pm 4$ \\
$E_{\mathrm{LO}}$ (meV) & $  / $ & $ 15 \pm 5$ \\
\bottomrule
\end{tabular}
\end{table}

\subsubsection{Polarization-dependent photoluminescence}

Polarization-resolved photoluminescence provides direct insight into the symmetry properties of the emitting electronic states and shows how intrinsic disorder manifests in the optical response of CuInSnS$_4$. As presented in Figure 7d, the BB emission exhibits an almost isotropic polarization response, with only weak angular variations within experimental uncertainty. Such behavior is expected for band-edge recombination in a crystal with an effectively isotropic electronic structure.

The DX emission shows a pronounced polarization anisotropy, characterized by a two-lobed angular modulation. Such behavior is indicative of electronic states with anisotropic transition dipole moments and is characteristic of localized excitons confined in locally symmetry-broken environments. Similar polarization anisotropies have been widely reported for excitonic emission in semiconductor quantum dots and other systems where local structural asymmetry or confinement lifts rotational symmetry and fixes the orientation of the excitonic transition dipole \cite{klenovsky2015polarization,yoon2021electrical,hermannstadter2012polarization}. In CuInSnS$_4$, this anisotropy is explained by intrinsic cation disorder, where asymmetric In/Sn configurations, distorted metal--sulfur coordination polyhedra, or short-range ordering motifs lower the local symmetry while remaining randomly distributed throughout the crystal.

This observed anisotropy reflects the internal symmetry of individual localized excitonic states rather than a macroscopic crystallographic axis. The absence of comparable anisotropy in the BB emission and in the vibrational response rules out extrinsic effects such as sample shape, optical alignment, or strain-induced birefringence as the dominant origin of the angular modulation. Instead, the polarization-dependent PL reveals that electronic excitations are uniquely sensitive to local symmetry variations that are largely invisible to diffraction and vibrational probes. Together with the power-dependent, temperature-dependent, and time-resolved PL results, these observations establish a consistent picture in which intrinsic cation disorder selectively localizes excitons and imprints directional character on their optical emission, while phonons continue to reflect the symmetry-averaged cubic lattice (as shown in Figure 1).

\begin{figure*}[t!]
    \centering
    \includegraphics[width=\textwidth]{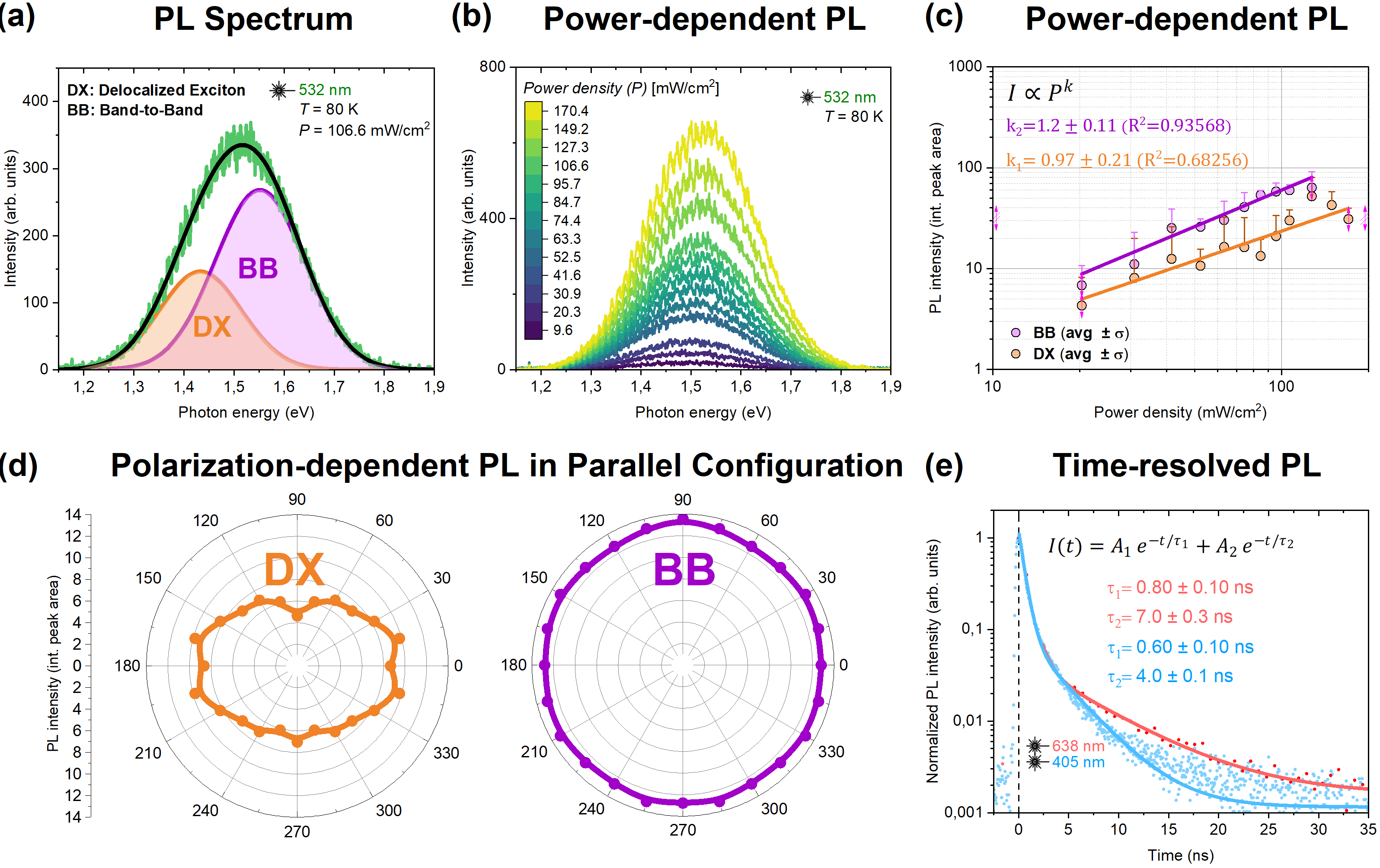}
    \caption{Optical, polarization and time-resolved \gls{pl} properties of \gls{cits}. 
(a) Representative \gls{pl} spectrum measured at \SI{80}{K} under \SI{532}{nm} excitation, fitted with two Gaussian components corresponding to band-to-band (BB) emission and a lower-energy disorder-related (DX) contribution. 
(b) Power-dependent \gls{pl} spectra recorded at \SI{80}{K} under \SI{532}{nm} excitation, showing the evolution of the emission profile with increasing excitation power density. 
(c) Log--log plot of the integrated \gls{pl} intensity versus excitation power density. Power-law fits yield exponents of $k = 1.2 \pm 0.11$ for the BB emission and $k = 0.97 \pm 0.21$ for the DX contribution, consistent with band-edge recombination and disorder-related recombination processes, respectively. 
(d) Polarization-dependent \gls{pl} intensity (integrated peak area) of the DX (orange) and BB (purple) emission components measured in parallel polarization configuration. The DX emission exhibits pronounced angular modulation, indicative of anisotropic disorder-related states, whereas the BB emission remains nearly isotropic. 
(e) Time-resolved \gls{pl} (\gls{trpl}) decays measured at \SI{295}{K} under \SI{405}{nm} and \SI{638}{nm} excitation. Bi-exponential fits reveal fast and slow recombination channels, with decay times of $\tau_1 = \SI{0.60}{ns} \pm \SI{0.1}{ns}$ and $\tau_2 = \SI{4.0}{ns} \pm \SI{0.1}{ns}$ for \SI{405}{nm}, and $\tau_1 = \SI{0.80}{ns} \pm \SI{0.10}{ns}$ and $\tau_2 = \SI{7.0}{ns} \pm \SI{0.3}{ns}$ for \SI{638}{nm} excitation.}

    \label{pl}
\end{figure*}

\subsubsection{Time-resolved PL and carrier dynamics}

TRPL measurements were performed to probe carrier recombination dynamics in CuInSnS$_4$. Representative decay transients recorded using 405~nm and 638~nm excitation at \SI{295}{K} are shown in Figure ~7e. In both cases, the decays are well described by a bi-exponential function,
\begin{equation}
I(t) = A_1 e^{-t/\tau_1} + A_2 e^{-t/\tau_2},
\label{eq:trpl}
\end{equation}
indicating the presence of two distinct recombination channels.

For 638~nm excitation, the extracted lifetimes are $\tau_1 = 0.80 \pm 0.10$~ns and $\tau_2 = 7.0 \pm 0.3$~ns, while excitation at 405~nm yields $\tau_1 = 0.60 \pm 0.10$~ns and $\tau_2 = 4.0 \pm 0.1$~ns. The fast sub-nanosecond component $\tau_1$ is assigned to rapid carrier trapping and non-radiative recombination from the shallow defects, surface states, or disorder-induced localized states. Such short decay times are significantly faster than typical radiative lifetimes in chalcogenide semiconductors and are characteristic of efficient carrier capture processes. The larger relative contribution of this fast component under higher-energy (405~nm) excitation is consistent with enhanced generation of hot carriers and increased sensitivity to near-surface and defect-rich regions.\cite{pean2020interpreting}

The slower component $\tau_2$, with lifetimes of several nanoseconds, is attributed to radiative recombination from band-edge and weakly localized band-tail states. The longer effective lifetime observed under lower-energy excitation reflects reduced excess carrier energy and a deeper excitation profile, which suppresses non-radiative losses and favors bulk radiative recombination. The coexistence of fast trapping-dominated and slower radiative decay channels is in agreement with the dual-emission behavior identified in steady-state PL, providing independent confirmation that carrier dynamics in CuInSnS$_4$ are governed by competition between disorder-assisted trapping and band-edge or weakly localized radiative recombination. These observations point to the importance of band-tail states and require a more quantitative assessment of disorder in this material.

\subsubsection{Quantification of disorder}

Disorder in CuInSnS$_4$ was quantified by extracting the Urbach energy $E_U$, which characterizes the exponential band-tail states arising from structural and thermal disorder. \cite{ugur2022life} The Urbach energy was obtained from the low energy tail of the PL spectra (Figure S5 in the Supporting Information), following the procedure of \citet{ugur2022life}. At low temperature (80~K), the relatively large Urbach energy $E_U \approx 47$~meV reflects substantial static disorder associated with structural distortions and cation disorder. With increasing temperature up to $\sim$200~K, $E_U$ decreases, consistent with thermal redistribution of carriers within band-tail states. At higher temperatures (240--280~K), $E_U$ increases again, indicating the growing contribution of dynamic, phonon-induced disorder. This crossover behavior supports the assignment of the low-energy PL emission to disorder-induced band-tail states and is consistent with the nanosecond-scale radiative lifetime observed in TRPL, where recombination proceeds through a thermally evolving distribution of localized states rather than discrete defect levels.

\subsection{Local vibrational and dielectric response measured using IR-SNOM}

To assess whether intrinsic cation disorder in \gls{cits} leaves detectable vibrational signatures on length scales inaccessible to far-field techniques, we employed \gls{snom}. Unlike conventional Raman spectroscopy, \gls{snom} offers nanometer-scale spatial resolution and is well suited to test whether local symmetry breaking associated with disorder produces nanoscale variations in the infrared dielectric response. The far-field and near-field spectra discussed here were acquired simultaneously at the same spatial location at multiple positions.

\begin{figure}[h]
    \centering
    \includegraphics[width=\columnwidth]{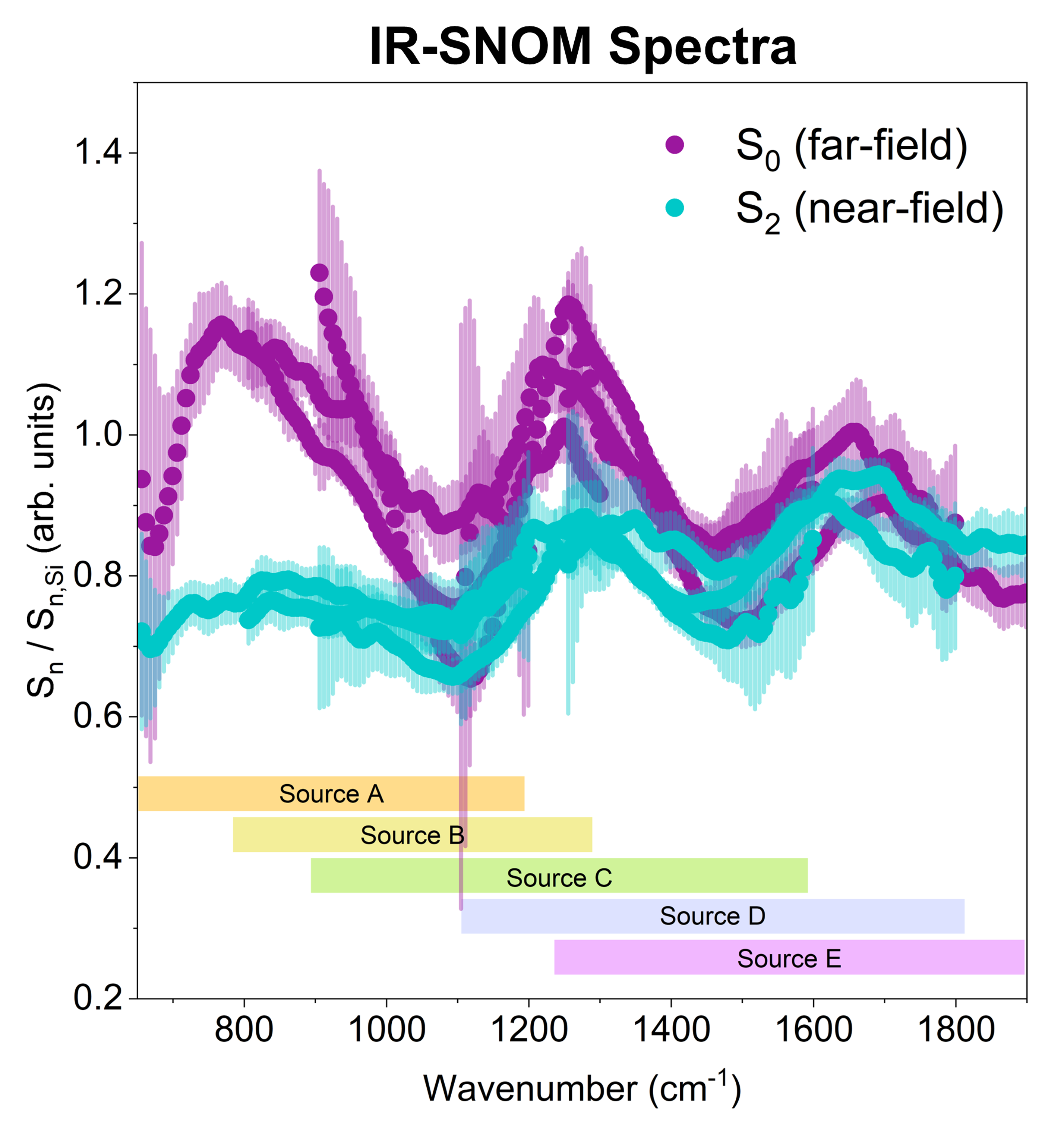}
    \caption{Comparison of the far-field-dominated ($n=0$) and near-field ($n=2$) spectra acquired from the same location on the CuInSnS$_4$ crystal and normalized to the corresponding spectra measured on a Si reference sample. The multiple traces arise from measurements performed with different infrared source channels, denoted as Sources A--E, each covering a distinct spectral window indicated by the colored horizontal bars. The vertical bars represent the experimental uncertainty of the measured amplitudes. Despite the use of multiple source channels, both the far-field and near-field spectra display a predominantly flat response with broadly similar spectral features across the investigated range.}
    \label{snom}
\end{figure}

Figure~\ref{snom} (a) compares the normalized scattering amplitude spectra measured in the far-field-dominated ($n=0$) and near-field ($n=2$) channels. Within experimental error, the two spectra show nearly identical spectral features and relative intensities across the investigated mid-infrared range. The relative difference in scattering amplitude remains low over most of the spectrum, with no systematic spectral shifts and no additional resonances in the near-field response.

The close agreement between the near-field and far-field spectra indicates that the infrared dielectric response of \gls{cits} remains homogeneous down to length scales of at least several tens of nanometers. Intrinsic In/Sn cation disorder therefore does not produce strongly localized infrared-active vibrational modes or pronounced dielectric heterogeneity. Instead, its effect on the vibrational response is limited to subtle mode mixing and linewidth broadening that remain collective in nature and are already captured by far-field infrared spectroscopy.

The local dielectric function shown in Figure~\ref{snom} (b) was extracted from the $n=2$ near-field spectra using the finite dipole model.\cite{cvitkovic2007analytical,hauer2012quasi} It provides information of the local complex permittivity, and the nanoscale infrared polarizability and absorption of CuInSnS$_4$. Both $\mathrm{Re}(\varepsilon)$ and $\mathrm{Im}(\varepsilon)$ vary smoothly across the investigated spectral range and do not reveal additional localized resonances or systematic anomalies beyond the weak variations already observed in the far-field response. This further supports the conclusion that the infrared dielectric response of CuInSnS$_4$ is spatially homogeneous on the length scale probed by s-SNOM, and that intrinsic In/Sn disorder remains averaged in the local dielectric response rather than giving rise to distinct nanoscale vibrational inhomogeneities.

\FloatBarrier

This behavior contrasts with the optical emission properties, where disorder gives rise to band-tail states and pronounced polarization anisotropy, highlighting that the electronic and excitonic responses in \gls{cits} are considerably more sensitive to local symmetry breaking than the vibrational dielectric function.

\section{Discussion}

Our combined vibrational and optical spectroscopy results show that intrinsic cation disorder in CuInSnS$_4$ manifests in a highly nonuniform manner. Lattice vibrations remain largely governed by the symmetry of the average cubic structure, despite the presence of local In/Sn disorder. Phonon energies, dispersions, and polarization responses are only weakly perturbed, indicating that the sulfur framework and short-range metal–sulfur bonding are preserved. This robustness reflects the close chemical similarity of In and Sn and the low energetic cost of cation mixing, which limits the impact of disorder on collective vibrational modes. As a result, phonons primarily probe an averaged structural potential, rather than local symmetry breaking.

In contrast, the electronic and excitonic responses reveal a strong sensitivity to local structural variations. PL measurements identify a disorder-related emission channel that is energetically close to the band edge but exhibits fundamentally different behavior from the band-to-band transition. Its temperature dependence, excitation-density response, and linewidth evolution are consistent with recombination from band-tail or weakly localized excitonic states rather than from isolated deep defects. This indicates that disorder reshapes the near-band-edge potential landscape without introducing strongly trapping centers, placing CuInSnS$_4$ in an intermediate disorder regime where localization occurs on short length scales while maintaining partial coupling to extended states.

Our key finding is that these localized electronic states carry a pronounced polarization anisotropy, whereas both the band-edge emission and the vibrational response remain effectively isotropic. This anisotropy cannot be attributed to macroscopic crystallographic effects or measurement geometry and instead reflects the intrinsic symmetry of individual localized excitonic states. Local arrangements of In and Sn ions distort the electrostatic and bonding environment on a scale relevant for electronic wavefunctions, fixing the orientation of the transition dipole moment. The absence of corresponding anisotropy in phonons or in the infrared dielectric response demonstrates that excitons are uniquely sensitive to symmetry breaking on length scales that are averaged out in lattice and dielectric measurements.

The ability to localize excitons and define polarization-selective optical transitions without relying on nanostructuring or introducing strong nonradiative losses enables new routes for engineering functional optical materials at the bulk or thin-film level. In multinary semiconductors such as CuInSnS$_4$, this approach can be exploited to realize polarization-addressable emitters, disorder-stabilized radiative recombination, and spatially confined excitonic states while maintaining vibrational and thermal robustness. Such characteristics are directly relevant for photonic devices requiring controlled emission properties and for quantum optical platforms where exciton localization and dipole orientation are essential. More broadly, the results demonstrate that chemically tunable disorder offers a scalable and energetically accessible strategy for tailoring optical functionality in complex semiconductors, positioning disorder engineering as a viable alternative to conventional structural confinement in next-generation optical and quantum materials.

\section{Conclusion}
We demonstrate that CuInSnS$_4$ exhibits a clear separation between vibrational and excitonic responses to intrinsic cation disorder. While lattice dynamics and dielectric properties reflect an averaged cubic symmetry, electronic excitations respond strongly to local symmetry breaking, giving rise to localized excitonic states with distinct polarization anisotropy and recombination dynamics. This phonon–exciton decoupling establishes CuInSnS$_4$ as a representative system in which disorder selectively influences electronic properties without strongly perturbing the lattice. Our results provide a potential framework for leveraging intrinsic cation disorder as a tool for engineering optical and excitonic functionality in multinary semiconductor materials.

\section{Experimental Section}
\subsection{\gls{cits} synthesis}
Crystal growth experiments were performed using the chemical vapor transport (CVT) technique. The starting metals Cu, In, and Sn together with the chalcogen sulfur were placed in a closed system, in this case a quartz glass ampoule with a diameter of 20~mm and a length of \SI{220}{mm}. The halogen iodine was used as a transport agent to enhance the material transport. 

The ampoule was evacuated, sealed, and placed in a two-zone furnace. At elevated temperatures iodine reacts with the elements to form volatile species. These halogen-containing species are transported from the hot zone to the cold zone of the ampoule by applying a temperature gradient $\Delta T$ across the growth tube. 

The transport process can be described by the following equilibrium reactions:

\begin{equation}
\begin{split}
\mathrm{Cu(s) + In(s) + Sn(s) + 4S(s) + 4I_2(s)} \\
\rightleftharpoons
\mathrm{CuI(g) + InI_3(g) + SnI_4(g) + 2S_2(g)}
\end{split}
\end{equation}

\begin{equation}
\mathrm{CuI(g) + InI_3(g) + SnI_4(g) + 2S_2(g) 
\rightleftharpoons CuInSnS_4(s) + 4I_2(s)}
\end{equation}

After cooling the furnace to ambient temperature, the ampoule was opened and the gaseous species present in the ampoule were allowed to evaporate.

For the growth of $\mathrm{CuInSnS_4}$ crystals, the metals and chalcogens were weighed to obtain 5~g of stoichiometric material. The concentration of the transport agent iodine was 5~mg\,cm$^{-3}$. The applied temperature gradient was $\Delta T = 100~^\circ$C, with a growth temperature of 700~$^\circ$C and a growth time of 240~hours.

\subsection{\gls{xrd} measurements}
Grown crystals were characterized by powder\gls{xrd} to determine the structural parameters. Synchrotron X-ray powder diffraction data were recorded at the KMC-2 beamline at BESSY~II\cite{tobbens2016kmc}, Helmholtz-Zentrum Berlin (HZB), using a wavelength of $\lambda = \SI{1.5406}{\angstrom}$ over a $2\theta$ range of \SIrange{5.8}{133.6}{\degree} with a step size of \SI{0.015}{\degree}.

The lattice parameters of the analyzed \gls{cits} crystal were determined by Le Bail analysis of the full diffraction pattern, applying the cubic spinel-type structure (space group $Fd\bar{3}m$) assuming In and Sn disordered on the octahedral position as the structural model in the refinement. The cubic lattice parameter was determined as 
\[
a = \SI{10.495}{\angstrom},
\]
with refinement quality parameters of $R_{\mathrm{Bragg}} = 2.55$ and $\chi^2 = 0.15$.

No indication of peak splitting was observed, which would be expected if In and Sn were ordered on two different Wyckoff positions, as in the distorted spinel structure with space group Imma.

\subsubsection{EDS measurements}
Compositional analyses were carried out using a Zeiss Gemini 450 scanning electron microscope (SEM) equipped with an energy-dispersive X-ray spectroscopy (EDS) detector. Prior to imaging, samples were mounted on aluminium SEM stubs using conductive carbon tape. The EDX spectra were acquired with an accelerating voltage of 10 kV to ensure sufficient excitation of characteristic X-ray lines. Elemental maps were generated by integrating the characteristic peaks of the relevant elements after background subtraction.

\subsection{Raman and photoluminescence measurements}

Raman and PL measurements were performed in a backscattering geometry using WITec alpha300 R and Horiba LabRAM confocal Raman microscopes. Multiwavelength Raman excitation was carried out using laser wavelengths of 325, 405, 488, 532, and 785~nm.  Depending on the excitation wavelength, the resulting laser spot size ranged between approximately 1 and 3~$\mu$m.

To avoid laser-induced heating or structural modification of the sample, power-dependent tests were performed prior to data acquisition. Raman spectra were recorded at a fixed position while systematically increasing the laser power density from the lowest available setting. The spectra were monitored for shifts in peak positions, linewidth broadening, or the appearance of additional features. The highest power density at which no such changes were observed was selected for all subsequent measurements.

The backscattered signal was analyzed using two spectrometer configurations. For excitation wavelengths of 325, 405, 488, and 532~nm, a spectrometer equipped with holographic gratings (150~g~mm$^{-1}$ for PL and 1800~g~mm$^{-1}$ for Raman measurements) and a thermoelectrically cooled CCD detector was used. Measurements at 785~nm were performed using a spectrometer fitted with a 1200~g~mm$^{-1}$ grating and a deep-depletion cooled CCD. All Raman spectra were acquired using a laser power density of 100~mW~cm$^{-2}$, with an integration time of 30~s and averaging over five accumulations to improve the signal-to-noise ratio. Raman peak positions were calibrated using the Si reference mode at 520~cm$^{-1}$. All room-temperature Raman measurements were conducted under ambient conditions. PL measurements were performed using 532~nm excitation with an integration time of 10~s, averaged over three accumulations.

Temperature-dependent Raman and photoluminescence measurements were performed using an Oxford Instruments Microstat He cryostat. The sample temperature was varied from 80 to 280~K and stabilized for 1 h prior to each acquisition to ensure thermal equilibrium. All spectra were recorded under identical optical alignment and excitation conditions to enable direct comparison across temperatures. 

\subsection{Time-resolved photoluminescence measurements}

TRPL measurements were performed using a MicroTime~100 time-resolved confocal microscope coupled to a PicoQuant detection unit. Photocarriers were excited using pulsed diode lasers at wavelengths of 405 and 638~nm, with pulse durations below 100~ps and a repetition rate of 20~MHz. The excitation beam was focused onto the sample surface through a long-working-distance 20$\times$ objective with a numerical aperture of 0.45.

The emitted photons were collected by the same objective and guided via a 50~$\mu$m core optical fiber to the detection unit, consisting of a FluoTime~300 photospectrometer equipped with a monochromator. The overall temporal resolution of the system was approximately 200~ps, which defines the uncertainty of the extracted decay times. All TRPL measurements were performed at \SI{295}{K}.

\subsection{IR-SNOM measurements}
The s-SNOM measurements were performed using a commercial system from neaSCOPE, Neaspec GmbH. The setup is equipped with a tunable broadband mid-infrared source from TOPTICA (600--2000~cm$^{-1}$) that provides illumination of the tip--sample region. An off-axis parabolic mirror focuses the IR radiation onto the sample. 

A Pt/Ir-coated nano-apex tip from NANOWORLD Arrow NCPt with a tip apex of $\sim$30~nm and an oscillation frequency of $\sim$255~kHz oscillates above the sample surface, providing near-field modulation. The backscattered light from the tip--sample region is collected by the same parabolic mirror and fed into an asymmetric interferometric detection configuration equipped with an MCT detector. The detected signal is sent to a lock-in amplifier to perform demodulation at higher harmonic orders of the tip tapping frequency.

A standard atomic force microcopy test sample consisting of SiO$_2$ islands on a Si substrate (TGQ1) was used to calibrate the system and provide reference spectra.

The dielectric function is extracted using the Finite dipole model\cite{cvitkovic2007analytical, hauer2012quasi}, where the near-field signal can be calculated from the sample dielectric function $\varepsilon$ according to

\begin{equation}
\frac{E_{\mathrm{out}}}{E_{\mathrm{in}}}
= (1+r)^2 \alpha_{\mathrm{eff}}
= (1+r)^2
\left(
1+\frac{1}{2}\frac{\beta f_0}{1-\beta f_1}+1
\right)
\end{equation}

where $r$ is the Fresnel reflection coefficient associated with $\varepsilon$, and 

\begin{equation}
\beta = \frac{\varepsilon-1}{\varepsilon+1}
\end{equation}

is the electrostatic reflection factor of the sample. The functions $f_{0,1}$ account for the geometry of the tip--sample coupling system and are given by

\begin{equation}
f_{0,1}
=
\left(
g-\frac{\rho + 2H + W_{0,1}}{2L}
\right)
\frac{
\ln\left(\frac{4L}{\rho + 4L + 2W_{0,1}}\right)
}{
\ln\left(\frac{4L}{\rho}\right)
}
\end{equation}

where $\rho = 30$~nm and $L = 300$~nm are the effective tip radius and tip length, respectively. The charge locations are approximated as $W_0 = 1.31\rho$ and $W_1 = 0.5\rho$. The factor 

\begin{equation}
g = 0.7 e^{0.06 i}\cite{hauer2012quasi}
\end{equation}

represents the proportion of induced charge responsible for near-field coupling.

Note that the tip oscillation introduces a sinusoidal modulation of the tip height. Consequently, the demodulated near-field signal at harmonic orders is calculated as the Fourier series at the $n$-th multiple of the tip oscillation frequency.

\subsection{Computational Methods}
Computational simulations were performed using Density Functional Theory (DFT) through the Vienna Ab Initio Simulation Package (VASP) \cite{kresse_initio_1993,kresse_initio_1994} with projector-augmented wave (PAW) pseudopotentials \cite{blochl_projector_1994}. 
A plane wave energy cutoff of \SI{350}{eV} and $k$-point density of \SI{0.26}{\angstrom^{-1}} ($4\times4\times4$ for the 14-atom Imma unit cell) were used, found to give total energies converged to within \SI{1}{meV/atom} using \texttt{vaspup2.0} \cite{kavanagh_vaspup20_2023}.
A 378-atom special quasirandom structure (SQS)\cite{zunger_special_1990} supercell was used to simulate the disordered cubic phase, generated using the Monte Carlo algorithm implemented in \texttt{icet}\cite{vandewalle_efficient_2013,angqvist_icet_2019,ekborg-tanner_construction_2024} with a $3\times3\times3$ expansion of the Imma unit cell.
The semi-local PBEsol DFT functional was used to relax this structure, with volume and cell shape as free parameters, and to compute force constants for the phonon dispersion.
The optical dielectric response was calculated using the method of Furthmüller et al.\cite{gajdos_linear_2006} to obtain the optical absorption spectrum.
PBEsol was also used for structural relaxation and phonon calculations for the ordered Imma phase.
To ensure accurate prediction of the electronic band gap, the range-separated screened hybrid DFT functional of Heyd, Scuseria and Ernzerhof (HSE06) \cite{heyd_hybrid_2003} was used, with the D3 correction \cite{grimme_consistent_2010} and zero-damping function, for atomic relaxation and calculation of the electronic structure for the ordered Imma phase.
The inclusion of spin-orbit coupling (SOC) was tested and found to have a negligible impact on the computed band gap ($\Delta E_{\textrm{g, SOC}} < \SI{0.01}{eV}$). A denser $k$-point grid of $8\times8\times8$ was used for the optical absorption calculation of the Imma phase.

\texttt{phonopy} \cite{togo_first_2015} and \texttt{ThermoParser} \cite{spooner_thermoparser_2024} were used to setup and parse the phonon calculations.
The 378-atom SQS supercell was directly used for phonon force constant calculations in the disordered cubic phase, followed by band unfolding to the primitive cubic cell using the \texttt{upho}\cite{ikeda_mode_2017} package, while a $4\times4\times4$ supercell of the primitive unit cell was used for Imma phonon calculations. 
\texttt{sumo}\cite{ganose_sumo_2018} was used to plot the calculated electronic band structure and absorption spectrum, \texttt{vasppy}\cite{morgan_vasppy_2021b} was used for radial distribution function analysis, and \texttt{pymatgen}\cite{ong_python_2013} was used throughout for manipulation and analysis of computational data.

\section*{Author contributions}
MD conceived, coordinated, and supervised the project. LKL performed the measurements, carried out the data analysis, produced the figures, and wrote the first draft of the manuscript. YT, GG, and DMT conducted the synthesis and XRD analysis. PA performed the compositional analysis. XL carried out the SNOM measurements and analysis. SK performed the DFT calculations and analysis. MC, RWC, JB, and SS contributed to supervision. All authors contributed to the interpretation of the results and provided feedback on the manuscript.

\section*{Conflicts of interest}
There are no conflicts to declare.

\section*{Data availability}
Data for this article, including XRD patterns, multiwavelength Raman spectra, polarization-dependent Raman spectra, temperature-dependent PL spectra, power-dependent PL spectra,  polarization-dependent PL spectra, time-resolved PL spectra, IR-SNOM spectra, extracted electric functions and the .cif files of the DFT calculations are available from Zenodo.org at https://doi.org/10.5281/zenodo.19114867.

\section*{Acknowledgment}
The authors acknowledge Angel Victor Labordet Alvarez and Guo Guanting for laboratory assistance. We thank ScopeM facilities for the Raman measurements with 325 nm excitation and HZB for the allocation of synchrotron radiation beamtimes.



\balance

\renewcommand{\refname}{References}
\bibliography{literature,SK_Zotero}
\bibliographystyle{rsc}

\newpage



\end{document}